\documentclass[longauth]{aa}

\usepackage{graphicx}
\usepackage{txfonts}
\usepackage{color}
\usepackage{soul}
\usepackage[normalem]{ulem}
\usepackage[rightcaption]{sidecap}
\usepackage{float}
\usepackage{xstring}
\usepackage{catchfile}
\newcommand{\gitfilename}{./.git/refs/heads/master.}
\IfFileExists{\gitfilename}{%
}

\begin{document}



%
%
  \title{Diurnal variation of dust and gas production in comet 67P/Churyumov-Gerasimenko at the inbound equinox as seen by OSIRIS and VIRTIS-M on board Rosetta}

   \author{C. Tubiana \inst{\ref{inst:MPS}}
                    \and G. Rinaldi\inst{\ref{inst:INAF_Rome}}
                    \and C. G\"uttler\inst{\ref{inst:MPS}}
                    \and C. Snodgrass\inst{\ref{inst:ROE}}
                    \and X. Shi\inst{\ref{inst:MPS}}
                    \and X. Hu\inst{\ref{inst:TUBerlin}}
					\and R. Marschall\inst{\ref{inst:ISSI}}
                    \and M. Fulle\inst{\ref{inst:OATrieste}}
                    \and D. Bockel\'ee-Morvan\inst{\ref{inst:LESIA}}
					\and G. Naletto\inst{\ref{inst:Astronomy_PD}}
					\and F. Capaccioni\inst{\ref{inst:INAF_Rome}}
                    \and H. Sierks\inst{\ref{inst:MPS}}
					\and G. Arnold\inst{\ref{inst:DLR_Berlin}}
					\and M. A. Barucci\inst{\ref{inst:LESIA}}
					\and J.-L. Bertaux\inst{\ref{inst:LATMOS}}
					\and I. Bertini\inst{\ref{inst:UNIPD}}
					\and D. Bodewits\inst{\ref{inst:Auburn_university}}
					\and M. T. Capria\inst{\ref{inst:INAF_Rome}}
					\and M. Ciarniello\inst{\ref{inst:INAF_Rome}}
					\and G. Cremonese\inst{\ref{inst:INAF_PD}}
					\and J. Crovisier\inst{\ref{inst:LESIA}}
					\and V. Da Deppo\inst{\ref{inst:CNR_PD}}
					\and S. Debei\inst{\ref{inst:DIE_PD}}
					\and M. De Cecco\inst{\ref{inst:UNI_TN}}
					\and J. Deller\inst{\ref{inst:MPS}}
					\and M.C. De Sanctis\inst{\ref{inst:INAF_Rome}}
					\and B. Davidsson\inst{\ref{inst:JPL}}
					\and L. Doose\inst{\ref{inst:LPL_Tucson}}
					\and S. Erard\inst{\ref{inst:LESIA}}
					\and G. Filacchione\inst{\ref{inst:INAF_Rome}}
          \and U. Fink\inst{\ref{inst:LPL_Tucson}}
					\and M. Formisano\inst{\ref{inst:INAF_Rome}}
					\and S. Fornasier\inst{\ref{inst:LESIA}}
					\and P. J. Guti{\'e}rrez\inst{\ref{inst:IAF}}
          \and W.-H. Ip\inst{\ref{inst:NCU},\ref{inst:Macao}}
					\and S. Ivanovski\inst{\ref{inst:OATrieste}}
					\and D. Kappel\inst{\ref{inst:Potsdam},\ref{inst:DLR_Berlin}}
					\and H. U. Keller\inst{\ref{inst:TU_BS},\ref{inst:DLR_Berlin}}
					\and L. Kolokolova\inst{\ref{inst:Univ_Maryland}}
					\and D. Koschny\inst{\ref{inst:ESTEC}}
					\and H. Krueger\inst{\ref{inst:MPS}}
					\and F. La Forgia\inst{\ref{inst:UNIPD}}
					\and P. L. Lamy\inst{\ref{inst:LATMOS_Guyancourt}}
				  \and L. M. Lara\inst{\ref{inst:IAF}}
					\and M. Lazzarin\inst{\ref{inst:UNIPD}}
					\and A. C. Levasseur-Regourd\inst{\ref{inst:LATMOS_2}}
					\and Z.-Y. Lin\inst{\ref{inst:NCU}}
					\and A. Longobardo\inst{\ref{inst:INAF_Rome},\ref{inst:DISP}}
					\and J. J. L{\'o}pez-Moreno\inst{\ref{inst:IAF}}
					\and F. Marzari\inst{\ref{inst:Astronomy_PD}}
					\and A. Migliorini\inst{\ref{inst:INAF_Rome}}
					\and S. Mottola \inst{\ref{inst:DLR_Berlin}}
					\and R. Rodrigo\inst{\ref{inst:INTA},\ref{inst:ISSI}}
					\and F. Taylor\inst{\ref{inst:Oxford}}
          \and I. Toth\inst{\ref{inst:Budapest}}
					\and V. Zakharov\inst{\ref{inst:INAF_Rome}}
         }
    \institute{Max Planck Institute for Solar System Research, G{\"o}ttingen, Germany\label{inst:MPS}\\   \email{tubiana@mps.mpg.de}
    \and Istituto di Astrofisica e Planetologia Spaziali, Istituto Nazionale di Astrofisica, Rome, Italy\label{inst:INAF_Rome}
    \and Institute for Astronomy, University of Edinburgh, Royal Observatory, Edinburgh EH9 3HJ, UK\label{inst:ROE}
    \and Institute for Geodesy and Geoinformation Science, Technical University Berlin, Stra{\ss}e des 17. Juni 135, 10623 Berlin, Germany\label{inst:TUBerlin}
    \and International Space Science Institute, Bern, Switzerland\label{inst:ISSI}
    \and INAF – Osservatorio Astronomico, Trieste, Italy\label{inst:OATrieste}
    \and LESIA, Observatoire de Paris, PSL Research University, CNRS, Sorbonne, Universit\'e, Univ.  Paris Diderot, Sorbonne Paris Cit\'e, 5 place Jules Janssen, F-92195 Meudon, France\label{inst:LESIA}
		\and Department of Physics, Oxford University, Oxford, UK\label{inst:Oxford}
		\and Deutsches Zentrum f\"ur Luft- und Raumfahrt (DLR), Institut f\"ur Planetenforschung, Berlin, Germany\label{inst:DLR_Berlin}
		\and Lunar and Planetary Laboratory, University of Arizona, Tucson, USA\label{inst:LPL_Tucson}
        \and Physics Department, Auburn University, Auburn, AL 36849, USA\label{inst:Auburn_university}
		\and Laboratoire Atmosph\`eres, Milieux et Observations Spatiales, CNRS \& Universit\'e de Versailles Saint-Quentin-en-Yvelines, 11 boulevard d'Alembert, 78280 Guyancourt, France\label{inst:LATMOS_Guyancourt}
    	\and Centro de Astrobiologia, CSIC-INTA, Torrejon de Ardoz, Madrid, Spain\label{inst:INTA}
		\and Science Support Office, European Space Research and Technology Centre/ESA, Keplerlaan 1, Postbus 299, 2201 AZ Noordwijk ZH, The Netherlands\label{inst:ESTEC}
		\and Jet Propulsion Laboratory, M/S 183-401, 4800 Oak Grove Drive, Pasadena, CA 91109, USA\label{inst:JPL}
		\and LATMOS, CNRS/UVSQ/IPSL, 11 Boulevard d'Alembert, 78280 Guyancourt, France\label{inst:LATMOS}
		\and University of Padova, Department of Physics and Astronomy ``Galileo Galilei'', Vicolo dell'Osservatorio 3, 35122 Padova, Italy\label{inst:UNIPD}
		\and INAF, Astronomical Observatory of Padova, Vicolo dell'Osservatorio 5, 35122 Padova, Italy\label{inst:INAF_PD}
		\and CNR-IFN UOS Padova LUXOR, Via Trasea 7, 35131 Padova, Italy\label{inst:CNR_PD}
		\and University of Padova, Department of Industrial Engineering, Via Venezia 1, 35131 Padova, Italy\label{inst:DIE_PD}
		\and University of Trento, Faculty of Engineering, Via Mesiano 77, 38121 Trento, Italy\label{inst:UNI_TN}
		\and Instituto  de Astrof\'{i}sica de Andaluc\'{i}a (CSIC), c/ Glorieta de la Astronomia s/n, 18008 Granada, Spain\label{inst:IAF}
		\and Institute of Physics and Astronomy, University of Potsdam, Potsdam, Germany\label{inst:Potsdam}
		\and Institut f\"ur Geophysik und extraterrestrische Physik, Technische Universit\"at Braunschweig, Mendelssohnstr. 3, 38106 Braunschweig, Germany\label{inst:TU_BS}
		\and University of Padova, Department of Physics and Astronomy ``Galileo Galilei'', Via Marzolo 8, 35131 Padova, Italy\label{inst:Astronomy_PD}
		\and Graduate Institute of Astronomy, National Central University, 300 Chung-Da Rd, Chung-Li 32054, Taiwan\label{inst:NCU}
		\and Space Science Institute, Macau University of Science and Technology, Avenida Wai Long, Taipa, Macau\label{inst:Macao}
		\and Department of Astronomy, University of Maryland, College Park, MD 20742-2421, USA\label{inst:Univ_Maryland}
		\and Konkoly Observatory, PO Box 67, 1525 Budapest, Hungary\label{inst:Budapest}
		\and LATMOS, Sorbonne Univ., CNRS, UVSQ, Campus Pierre et Marie Curie, BC 102, 4 place Jussieu, 75005 Paris, France\label{inst:LATMOS_2}
		\and DIST, Universita Parthenope, Centro Direzionale Isola C4, 80100 Napoli, Italy\label{inst:DISP}
		}
%
%
 %
  \abstract
  {On 27 April 2015, when 67P/Churyumov-Gerasimenko was at 1.76 au from the Sun and moving towards perihelion, the OSIRIS and VIRTIS-M instruments on board Rosetta simultaneously observed the evolving dust and gas coma during a complete rotation of the comet.}
  {We aim to characterize the dust, H$_2$O and CO$_2$ gas spatial distribution in the inner coma. 
  To do this we performed a quantitative analysis of the release of dust and gas and compared the observed H$_2$O production rate with the one calculated using a thermo-physical model.}
   {For this study we selected OSIRIS WAC images at 612 nm (dust) and VIRTIS-M image cubes at 612 nm, 2700 nm (H$_2$O emission band) and 4200 nm (CO$_2$ emission band).
We measured the average signal in a circular annulus, to study spatial variation around the comet, and in a sector of the annulus, to study temporal variation in the sunward direction with comet rotation, both at a fixed distance of 3.1 km from the comet centre.}
   {The spatial correlation between dust and water, both coming from the sun-lit side of the comet, shows that water is the main driver of dust activity in this time period. The spatial distribution of CO$_2$ is not correlated with water and dust.
There is no strong temporal correlation between the dust brightness and water production rate as the comet rotates. The dust brightness shows a peak at 0\degr\ sub-solar longitude, which is not pronounced in the water production. At the same epoch, there is also a maximum in CO$_2$ production.
An excess of measured water production, with respect to the value calculated using a simple thermo-physical model, is observed when the head lobe and regions of the Southern hemisphere with strong seasonal variations are illuminated (sub-solar longitude 270\degr -- 50\degr).
A drastic decrease in dust production, when the water production (both measured and from the model) displays a maximum, happens when typical Northern consolidated regions are illuminated and the Southern hemisphere regions with strong seasonal variations are instead in shadow (sub-solar longitude 50\degr -- 90\degr).
Possible explanations of these observations are presented and discussed.
}
   {}
   \titlerunning{The properties of dust and gas in the coma of 67P}
   \maketitle
%

\section{Introduction}
\label{sec:introduction}
After 10 years of cruise and 30 months of deep space hibernation, the ESA {\it{Rosetta}} spacecraft woke up on 20 January 2014. 
Rosetta had the unique opportunity to stay in the vicinity of comet 67P/Churyumov-Gerasimenko (hereafter 67P) for 2.5 years, observing how the comet evolved while moving along its orbit. 

One of the main goals of Rosetta is to understand cometary activity, i.e. the physical processes generating the dust and gas coma from the nucleus.
While the broad picture of the Whipple model, i.e., ices in the nucleus sublimate when heated by the Sun and the resulting gas outflow lifts dust \citep{Whipple}, has been confirmed by observations, details of the processes involved remain the subject of debate. Observations of dust `jets' have been traced to certain areas of the surface \citep{OSIRIS_MPS_2016_Vincent_MNRAS}, and various models have been developed to trace gas and dust flow in the inner coma of 67P, with varying degrees of complexity \citep[e.g.,][]{Fougere2016,Kramer2016,Kramer2017,Zakharov2018}, but these do not yet uniquely identify the surface features responsible for activity. Indeed, many of the models show that the bulk activity can be explained by more-or-less homogeneous activity from all illuminated surface facets \citep{Keller2015}, and that jets in the inner coma are controlled more by the complex shape of the nucleus than by anything special about their apparent source on the surface \citep{Shi2018}. Yet, there are clear variations in activity with seasonal illumination of the comet, 
which appear to be related to the very different morphology of the Northern and Southern surfaces, 
and models that explain the early activity seen by Rosetta when 67P was far from the Sun do not reproduce the perihelion behaviour \citep{ShiEtAl2018EPSC}. 
Recent models attempt to reproduce the complexities of the changing activity \citep{Attree2019,Marschall2019Icarus}, including investigating the relative contribution of different sublimating ices (water or CO$_2$) to driving activity, i.e., how they are related to each other \citep{Gasc2017} and to dust release.

The aim of this work is to take advantage of the capabilities of two instruments on Rosetta to analyse the dust and gas coma behaviour in the pre-perihelion phase, when the comet was at heliocentric distances of 1.76 au and close to the equinox between the changing seasons, allowing us to investigate the differences in observed dust and gas distributions  and their longitudinal variations simultaneously over a full rotation of the nucleus. We investigate both the spatial variation in dust and gas around the comet, and how it varies with time as different areas are illuminated throughout the comet day.

\section{Instruments, datasets and methods}
\label{sec:dataset_methods}
\subsection{Instruments}
\label{sec:instruments}

The Optical, Spectroscopic, and Infrared Remote Imaging System (OSIRIS) and the Visible InfraRed Thermal Imaging Spectrometer (VIRTIS) are two of the 12 scientific instruments on the Rosetta orbiter \citep{2007_Glassmeier_SSR}. 
OSIRIS \citep{OSIRIS_2007_Keller_SSR} is the scientific camera system. 
It comprises a Narrow Angle Camera (NAC) and a Wide Angle Camera (WAC) with a field of view (FOV) of 2.20$\degr$ $\times$ 2.22$\degr$ and 11.35$\degr$ $\times$ 12.11$\degr$, respectively.
Both cameras use a 2048 $\times$ 2048 pixel backside illuminated CCD detector with a UV optimized anti-reflection coating. 
The CCDs are equipped with lateral anti-blooming that allows overexposure of the nucleus without creating saturation artifacts, enabling the study of details in the faint coma structures next to the illuminated limb.
The NAC is equipped with 11 filters covering the wavelength range 250 -- 1000 nm, while the WAC has 14 filters covering the range 240 -- 720 nm \citep{OSIRIS_MPS_2015b_Tubiana_A&A}.

The VIRTIS spectrometer\citep{Coradini2007} is composed of two spectral channels: VIRTIS-M and VIRTIS-H. 
VIRTIS-M is the visible (230 -- 1000 nm, 432 bands) and infrared (1000 -- 5000 nm, 432 bands) imaging spectrometer with a field of view of 3.6\degr\ (along the slit axis) and an instantaneous field of view (IFOV) of 250 $\mu$rad. 
The instrument acquires hyperspectral cubes by scanning in time the target scene line by line. 
The duration of the acquisition ($\Delta t$ in Table \ref{tab:VIRTISobservationalDetails}) is given by the number of lines (including periodic dark current frames) times the internal repetition time, where the repetition time is the time between two consecutive steps necessary to move the internal scan mirror by one IFOV. 
The integration time ($t_{exp}$  in Table \ref{tab:VIRTISobservationalDetails}) set for the VIS and IR channels is lower than the internal repetition time. 
The maximum 3.6\degr\ $\times$ 3.6\degr\ FOV is imaged by repeating acquisition on successive 256 scan mirror steps (lines). 
From a distance of 100 km this corresponds to a 6.4 km $\times$ 6.4 km swath with a resolution of 25 m/pix. 
As an example, we show in Fig. 
\ref{fig:virtis_image_time} 
a VIRTIS-M hyperspectral cube with the line and time axes. 
VIRTIS-H is the infrared high-spectral resolution point spectrograph operating in the 1900 -- 5000 nm spectral range with a $ \lambda/ \Delta \lambda = 1300 \div 3000$. 
The instrument observes in a single IFOV of 580 $\mu$rad $\times$ 1740 $\mu$rad, which corresponds to a resolution of 58 m $\times$ 174 m from a 100 km distance. 
Since the VIRTIS-H and VIRTIS-M boresights are co-aligned, during the time necessary for a scan for the imaging channel the point spectrograph can acquire the same area multiple times. 

\subsection{Datasets}
\label{sec:dataset}

\begin{table*}[t]
\caption{OSIRIS observational details.}
\label{tab:OSIRISobservationalDetails}
\footnotesize\centering
\begin{tabular}{llccccc}
\hline
\hline
\\
\# &Filename & $t_\text{start}$ [UTC] & $t_\text{exp}$ [s] & $d_\text{S/C}$ [km] & $R$ [m$/$pix] &$\phi$ [\degr] \\
\\
\hline
\hline
\\
a&WAC$\_$2015-04-27T09.24.16.464Z$\_$ID30$\_$1397549300$\_$F18.IMG & 09:25:32 & 7.8& 125& 12.4 &249.7\\
b&WAC$\_$2015-04-27T10.24.16.524Z$\_$ID30$\_$1397549400$\_$F18.IMG & 10:25:32 & 7.8& 126& 12.6&220.7
\\
c&WAC$\_$2015-04-27T11.25.18.671Z$\_$ID30$\_$1397549500$\_$F18.IMG & 11:26:34 & 7.8& 127& 12.7& 191.3
\\
d&WAC$\_$2015-04-27T13.04.01.065Z$\_$ID30$\_$1397549200$\_$F18.IMG & 13:05:16 & 7.8& 129& 12.9 &143.6
\\
e&WAC$\_$2015-04-27T13.59.00.784Z$\_$ID30$\_$1397549300$\_$F18.IMG & 14:00:16 & 7.8& 130& 13.0& 117.1 
 \\
f&WAC$\_$2015-04-27T14.59.00.701Z$\_$ID30$\_$1397549400$\_$F18.IMG & 15:00:16 & 7.8& 132& 13.1&   88.1
\\
g&WAC$\_$2015-04-27T15.59.00.505Z$\_$ID30$\_$1397549500$\_$F18.IMG & 16:00:16 & 7.8& 133& 13.2& 59.2
\\
h&WAC$\_$2015-04-27T16.28.59.503Z$\_$ID30$\_$1397549400$\_$F18.IMG & 16:30:15 & 7.8& 134& 13.3&44.7
 \\
i&WAC$\_$2015-04-27T17.29.57.713Z$\_$ID30$\_$1397549100$\_$F18.IMG & 17:31:13 & 7.8& 135& 13.4 &15.3 
\\
j&WAC$\_$2015-04-27T18.17.57.683Z$\_$ID30$\_$1397549200$\_$F18.IMG & 18:19:13 & 7.8& 136& 13.5 & 352.1
\\
k&WAC$\_$2015-04-27T19.17.57.677Z$\_$ID30$\_$1397549300$\_$F18.IMG & 19:19:13 & 7.8& 137& 13.7 & 323.2
\\
l&WAC$\_$2015-04-27T20.17.57.647Z$\_$ID30$\_$1397549400$\_$F18.IMG & 20:19:13 & 7.8& 139& 13.8& 294.2\\
m&WAC$\_$2015-04-27T20.42.57.691Z$\_$ID30$\_$1397549300$\_$F18.IMG & 20:44:13 & 7.8& 139& 13.9 &282.2
\\
n&WAC$\_$2015-04-27T22.33.00.788Z$\_$ID30$\_$1397549400$\_$F18.IMG & 22:34:16 & 7.8& 142& 14.1 &229.1\\
\hline
\end{tabular}
\\
\flushleft
{\bf{Note}}:
{\it{Column 1}}: Assigned letter for each image.
{\it{Column 2}}: Observations filenames.
{\it{Column 3}}: Start time of each image cube [UTC].
{\it{Column 4}}: Exposure time.
{\it{Column 5}}: S/C distance from the comet centre.
{\it{Column 6}}: Pixel dimension at the distance of each observation.
{\it{Column 7}}: Sub-solar longitude of each observation.
\end{table*}

\begin{table*}[t]
\caption{VIRTIS-M observational details.}
\label{tab:VIRTISobservationalDetails}
\footnotesize\centering
\begin{tabular}{llcccccccc}
\hline
\hline
\\
\# & Filename & $S_\text{cube}$ & $t_\text{start}$ [UTC] & $\Delta t$ [s]& $t_\text{exp}$ [s] & $d_\text{S/C}$ [km]& $R$ [m/pix]& $\phi_{\pm45\degr}$ [\degr]& $\phi_{360\degr}$ [\degr]\\
\\
\hline
\hline
\\
&VIRTIS-M-IR \\
\hline
\\
1&I1$\_$00388760487.CAL &256 264 432& 13:02:43 & 2778 &3 &128& 32&140.0 &132.9   
   \\
2&I1$\_$00388763546.CAL & 256 264 432&13:53:41 & 2778& 3 &130&32&115.4 &108.3 
\\
3&I1$\_$00388766847.CAL & 256 264 432& 14:48:43 & 2778& 3&131&33&88.7  &81.7   
\\
4&I1$\_$00388770446.CAL & 256 264 432& 15:48:41 & 2778& 3&132&33&59.7  &52.8  
\\
5&I1$\_$00388776027.CAL &256 258 432 & 17:21:43 & 5415&3&134&34&9.3 &355.9   
\\
6&I1$\_$00388781546.CAL & 256 258 432& 18:53:41 & 5415&3&136&34&324.8 &311.6  
\\
7&I1$\_$00388787067.CAL & 256 220 432& 20:25:43 & 4635&3&139&35&280.2  &268.4  
\\
8&I1$\_$00388794147.CAL & 256 133 432& 22:23:43 & 1398& 3&141&35&228.6 &222.9\\
9&I1$\_$00388795646.CAL & 256 133 432& 22:48:41 & 1398& 3&142&35&216.6& --\\
\hline
\\
&VIRTIS-M-VIS \\
\hline
\\
10&V1$\_$00388760489.CAL &256 264 432& 13:02:45 & 2778 & 5&128& 32&140.0&132.9 \\
11&V1$\_$00388763549.CAL & 256 264 432&13:53:45 & 2778&5&130&32&115.4 &108.3\\
12&V1$\_$00388766849.CAL & 256 264 432& 14:48:45 & 2778&5&131&33&88.7&81.7\\
13&V1$\_$00388770449.CAL & 256 264 432& 15:48:45 & 2778&5&132&33&59.7&52.8\\
14&V1$\_$00388776036.CAL &256 258 432 & 17:21:52 & 5415&5&134&34&9.3&355.9 \\
15&V1$\_$00388781556.CAL & 256 258 432& 18:53:52 & 5415&5&136&34&324.8 &311.6\\
16&V1$\_$00388787076.CAL & 256 220 432& 20:25:52 & 4635&5&139&35&280.2&268.4  \\
17&V1$\_$00388794149.CAL & 256 133 432& 22:23:45 & 1398&5&141&35&228.6&222.9\\
18&V1$\_$00388795649.CAL & 256 133 432& 22:48:45 & 1398&5&142&35&216.6& --\\
\hline
\end{tabular}
\flushleft
{\bf{Note}}:
{\it{Column 1}}: Assigned number for each image cube.
{\it{Column 2}}: Observations filenames.
{\it{Column 3}}: Image cube dimension in number of samples (256 fixed pixel number), number of scan lines and spectral bands (432 for each channel).
{\it{Column 4}}: Start time of each image cube [UTC].
{\it{Column 5}}: Total duration for each image cube.
{\it{Column 6}}: Exposure time for each line.
{\it{Column 7}}: S/C distance from the comet centre.
{\it{Column 8}}: Pixel dimension at the distance of each observation.
{\it{Column 9}}: Sub-solar longitude of each observation at the time of the middle of the $\pm$ 45$^{\circ}$ sector (see Fig. \ref{fig:virtis_image_time} and Sec. \ref{sec:method}).
{\it{Column 10}}: Sub-solar longitude of each observation at the time of the middle of the annulus (see Fig. \ref{fig:virtis_image_time} and Se. \ref{sec:method}).
\end{table*}

Throughout the entire mission, we regularly carried out so-called {\it{dust monitoring}} sequences, which are designed to observe the coma of unresolved dust particles.
Typically, these sequences span at least 12 hours (i.e. a full comet rotation) with hourly cadence. 
In the frame of this work, we have analysed one dust monitoring sequence that was acquired on 27 April 2015\footnote{The data are available at the Planetary Science Archive of the European Space Agency under https://www.cosmos.esa.int/web/psa/rosetta}. 
At the time of the observation, 67P was at a heliocentric distance of 1.76 au, moving towards perihelion, and close to the equinox between the long but cool Northern summer and the short, intensely illuminated Southern summer around perihelion. The Rosetta spacecraft was at a distance between 125 and 142 km from the comet.
The Rosetta +Z-axis was pointing to the comet nucleus ({\it{IlluminatedPoint}} pointing) and, given the spacecraft-comet distance, the nucleus was entirely contained in all OSIRIS WAC images and in 7 out of 9 VIRTIS-M images.

The spacecraft was approximately in a terminator orbit (phase angle $\sim$ 73\degr\ -- 75\degr) so that one side of the comet was illuminated by the Sun and the other side was in darkness. 
The observational details are summarised in Table \ref{tab:OSIRISobservationalDetails} and Table \ref{tab:VIRTISobservationalDetails}.
The 27 April 2015 
dataset is one of the best acquired by VIRTIS-M for this purpose, as it covers both the VIS and IR channels and spans more than 12 hours. It was one of the final monitoring observations before the failure of the VIRTIS-M cryocooler disabled the IR channel. 
It was optimised for coma observations, achieving a high signal-to-noise ratio. 
The same data set was previously analysed by \citet{Rinaldi2016} and \citet{Fink2016}.
\citet{Rinaldi2016} focused their analysis on the comparison between dust and H$_2$O and CO$_2$ gas  spatial distributions, radial profiles and azimuthal distributions to search for any correlation between them.
\citet{Fink2016} focused their investigation on the emission intensity of CO$_2$ and H$_2$O and provided an explanation for the large observed variations reported in the literature for the CO$_2$/H$_2$O ratio. 
In this work we study temporal variation in more detail, and can better study areas of the dust continuum badly affected by stray light in the VIRTIS-M data, by combining it with OSIRIS imaging.

\subsubsection{OSIRIS dataset}
\label{sec:dataset_osiris}

\begin{SCfigure*}
	\centering
	\includegraphics[width=0.5\textwidth,trim=0cm 0.0cm 0.5cm 0.0cm,clip=true]{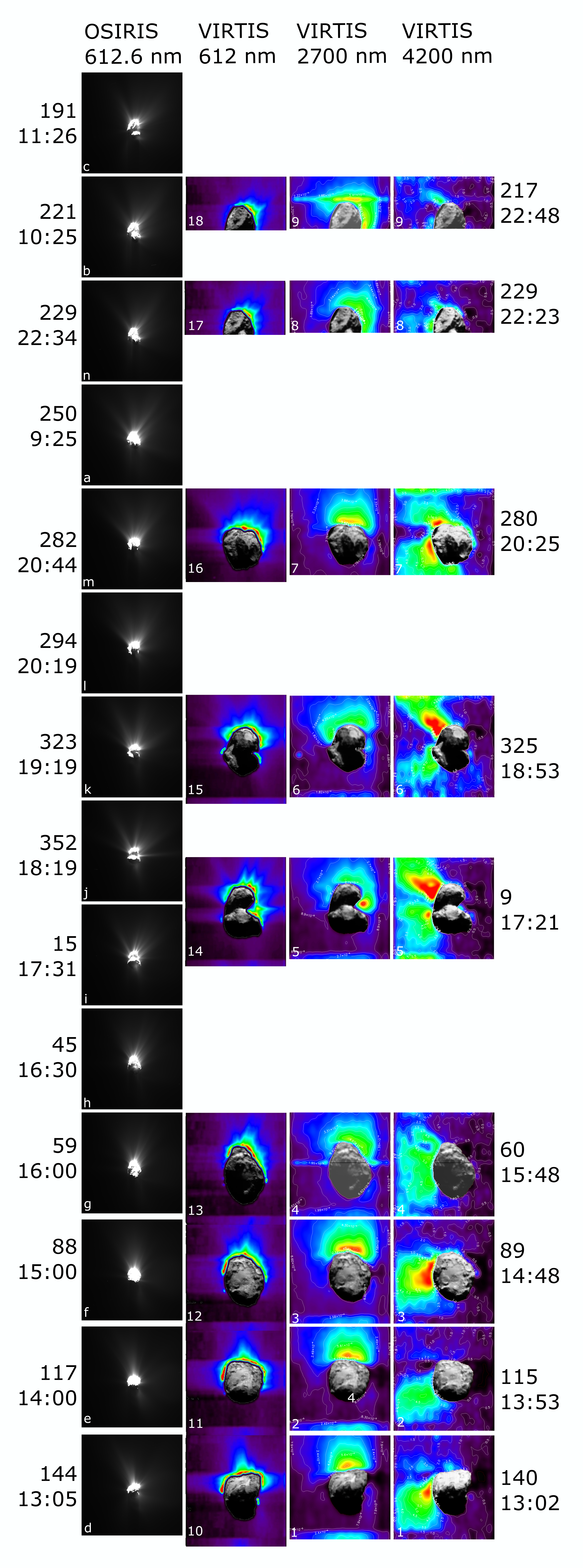}
	\caption{OSIRIS and VIRTIS-M images used for the analysis. 
		For each dataset, the images are scaled to the same brightness level and are displayed in the standard Rosetta orientation, with the Sun up.
		{\it{First column}}: OSIRIS WAC images in the VIS610 filter. 
		On the left hand side, the sub-solar longitude (in \degr) and start time of each image (in UTC) are listed. 
		{\it{Second column}}: VIRTIS-M image at 612 nm. 
		{\it{Third and fourth columns}}: Band intensity maps of H$_2$O and CO$_2$, respectively.
		For better visualisation, a VIRTIS-M image at 4200 nm was inserted into the nucleus area in the images at 612 nm, 2700 nm and 4200 nm.  
		On the right hand side, the sub-solar longitude (in \degr) and start time of each image (in UTC) are listed.
		Each image is labelled with the assigned number listed in column 1 of Table \ref{tab:OSIRISobservationalDetails} and Table \ref{tab:VIRTISobservationalDetails}.}
	\label{fig:images_osiris_virtis}
\end{SCfigure*}

The dust monitoring sequence STP053\_DUST\_MON\_006 contained 45 WAC full-frame images. 
For this study we have selected the 15 images acquired with the VIS610 filter ($\lambda_{cent}$ = 612.6 nm, $\Delta\lambda$ = 9.8 nm) and with exposure time optimized for dust coma studies. 
One image was acquired with the WAC door closed and excluded from the analysis. 
Thus, the total number of OSIRIS images used is 14. 
We used OSIRIS level 3 (CODMAC Level 4) images, which are radiometric calibrated and geometric distortion corrected (for details see a description of the OSIRIS calibration pipeline in \citet{OSIRIS_MPS_2015b_Tubiana_A&A}). 
The images, scaled to the same intensity levels, are shown in Fig. \ref{fig:images_osiris_virtis} (first column).

\subsubsection{VIRTIS-M dataset}
\label{sec:dataset_virits}

\begin{figure}[t]
\centering
\includegraphics[width=0.8\columnwidth,trim=0.7cm 0.1cm 1cm 1cm,clip=true]{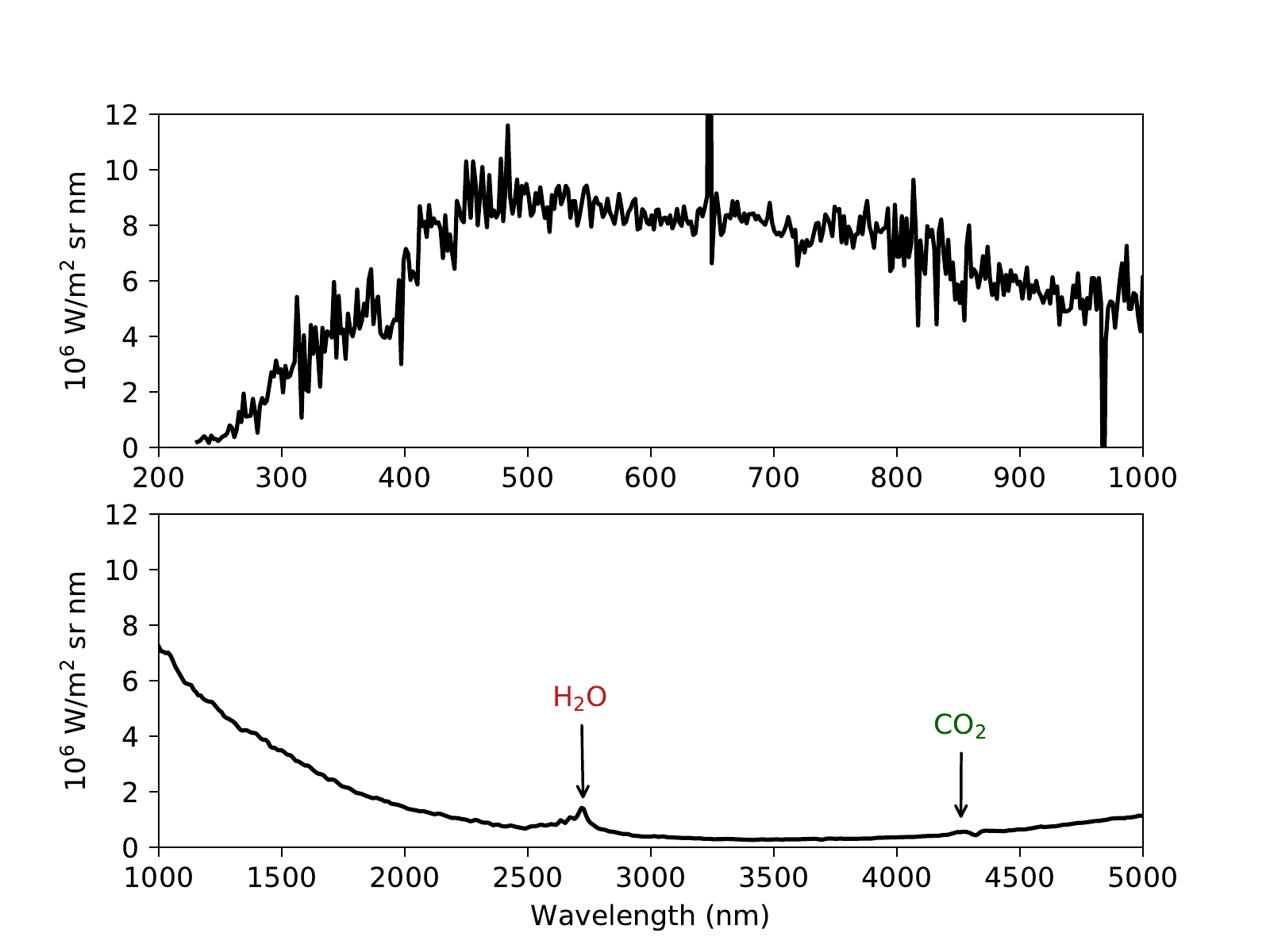}
\caption{Radiance spectrum for a VIRTIS-M cube in the VIS channel (V1\_00388760489) (upper plot) and in the IR channel (I1\_00388760487) (lower plot).  
The dust continuum, in the range 200 -- 3000 nm, is due to Sun light scattered by the dust particles in the coma. 
The IR spectrum shows the gas fluorescence emission of water vapour and CO$_2$ at 2700 nm and 4200 nm, respectively.}
\label{fig:virtis_spectra}
\end{figure}

A set of nine image cubes was obtained.
The image cubes in the two channels were taken at the same time, with the VIS exposure sequences starting about 2-4 s after the IR ones. 
The first seven cubes have a FOV of about 9.0 km $\times$ 7.7 km (at the nucleus centre distance) with the comet nucleus position roughly in the centre of the image.
The last two cubes have a FOV of about 9.0 km $\times$ 4 km and the nucleus is only partially contained in the frame.
Figure \ref{fig:images_osiris_virtis} shows the spatial distribution of dust (612 nm), water vapour (2700 nm) and carbon dioxide (4200 nm).
Each map is a composite image where the comet nucleus, taken at 4200 nm, is superimposed on the maps of the the dust continuum at 612 nm (second column), the water vapour (third column) and the CO$_2$ band intensities (fourth column). 
When the bright nucleus partially illuminates the instrument's slit a sizeable fraction of the incoming photons are spread into the adjacent coma pixels. 
The measured dust continuum is contaminated by nucleus stray light and cannot be used.
The data cubes used for the analysis were calibrated using the VIRTIS reduction pipeline as described by \citet{Ammannito2006}, \citet{Filacchione2006} and \citet{Rinaldi2016}. 

A typical radiance spectrum of the 67P coma in the VIS and IR is shown in Fig. \ref{fig:virtis_spectra}. 
It demonstrates the unique capability of the VIRTIS-M instrument to simultaneously measure the dust continuum in the range 200 -- 3000 nm and the fluorescence emission of water vapour and CO$_2$. The calculation of H$_2$O and CO$_2$ band intensity is described by \citet{Migliorini2016} and \citet{Fink2016}.

To measure the dust continuum intensity, we chose a  9.8 nm wide band centred at 612.6 nm, which corresponds to the OSIRIS WAC VIS610 filter. 
This allowed a comparison of the results obtained by the two instruments.

\subsection{Aperture photometry}
\label{sec:method}

\begin{figure}
\centering
\includegraphics[width=1.0\columnwidth,trim=0.9cm 2.0cm 0cm 0cm,clip=true]{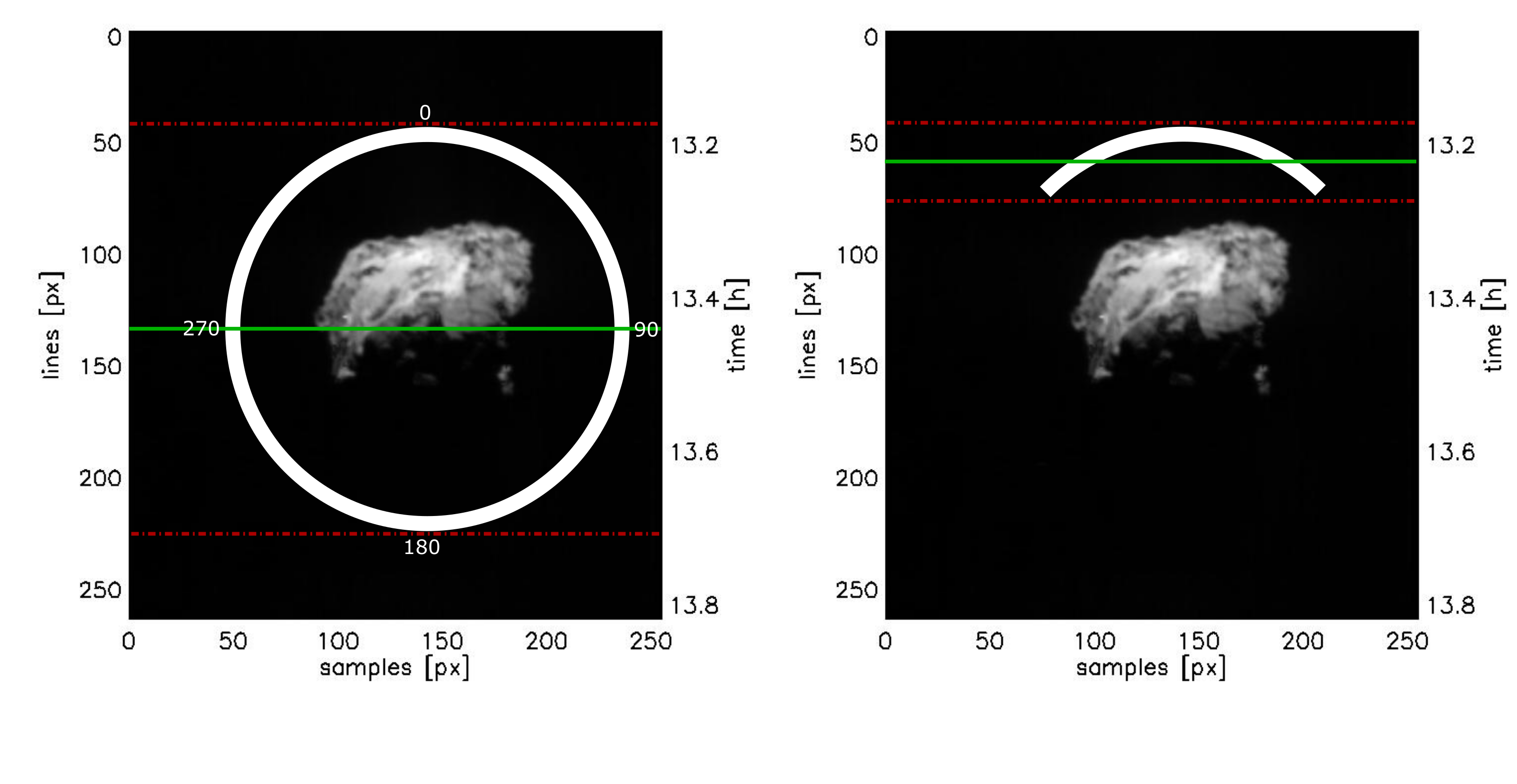}

\caption{Example of masks used for aperture photometry measurements: the $\pm$ 45$^{\circ}$ sector (on the right) and the annulus (on the left) superimposed on a VIRTIS-M image at 1100 nm (selected to show the nucleus at the same scale). The two vertical axes on each image show the operation mode of VIRTIS-M: the instrument scans spatially (number of line on the left) and temporally (time on the right) through the coma and nucleus of the comet. The green line indicates the mid-point inside each mask for the $\pm$ 45$^{\circ}$ sector and the full annulus.
}
\label{fig:virtis_image_time}
\end{figure}

In each image, we have measured the average signal in an annulus, or in a sector of the annulus, at fixed distance (in km) from the centre of the comet (Fig. \ref{fig:virtis_image_time}). 
We call {\it{mask}} the selected area where the flux is measured.
We chose a mask {\it{width}} of 0.2 km in radial direction for the entire dataset.
As {\it{distance}} we have chosen the maximum distance from the comet centre for which the mask is fully included in the VIRTIS-M frame, which is smaller than the OSIRIS WAC FOV. 
For this dataset the selected distance is 3.1 km.
Due to the irregular shape of the comet, a single-circle mask with a fixed distance from the centre of the comet is not equidistant to the comet's limb.

Due to the VIRTIS-M operation mode (see Sec. \ref{sec:instruments}) each image cube line was acquired at a different time and therefore has a different longitude of the sub-solar point.
For each image cube, we have determined the mid-point inside each mask, as shown in Fig. \ref{fig:virtis_image_time}, and calculated the sub-solar longitude of the mid-point. 
Since the sub-solar longitude is mask-dependent because the number of considered lines is different in the two masks, the sub-solar longitude is different for a sector or for an annulus in the same image cube (see Table \ref{tab:VIRTISobservationalDetails}).

For the OSIRIS images, the statistical error associated to each measurement is determined using the sigma (or error) map \citep{OSIRIS_MPS_2015b_Tubiana_A&A}. 
It contains the error associated with the intensity of each pixel, calculated using Poisson statistics and the readout noise error. 
This statistical error is very small, of the order of $\pm$ 0.3\%. 
In addition to this statistical error, the images have a systematic error, due to the radiometric calibration, of $\pm$ 1\% \citep{OSIRIS_MPS_2015b_Tubiana_A&A}. 
VIRTIS-M measurements have a statistical error given by the standard deviation of the average flux inside the mask and a systematic error due to the radiometric calibration. 
The total uncertainty on the measurement, calculated using error propagation, is of the order of $\pm$ 10\% \citep{Filacchione2006, Coradini2007}.

\subsection{Azimuthal profiles}
\label{sec:azimuthal_prof}
To determine the azimuthal profiles of dust and gas we used the circular mask at 3.1 $\pm$ 0.2 km from the centre of the comet. The selected angular step is 10\degr. The profiles are measured clockwise. 0\degr\ is in the sub-solar direction, as sketched in Fig. \ref{fig:virtis_image_time}.

\section{Dust coma at 612 nm}
\subsection{Longitudinal variation}
\label{sec:dust_diurnal_variation}

\begin{figure}[t]
\centering
\includegraphics[width=1.0\columnwidth,trim=0.5cm 0cm 1.0cm 1.0cm,clip=true]{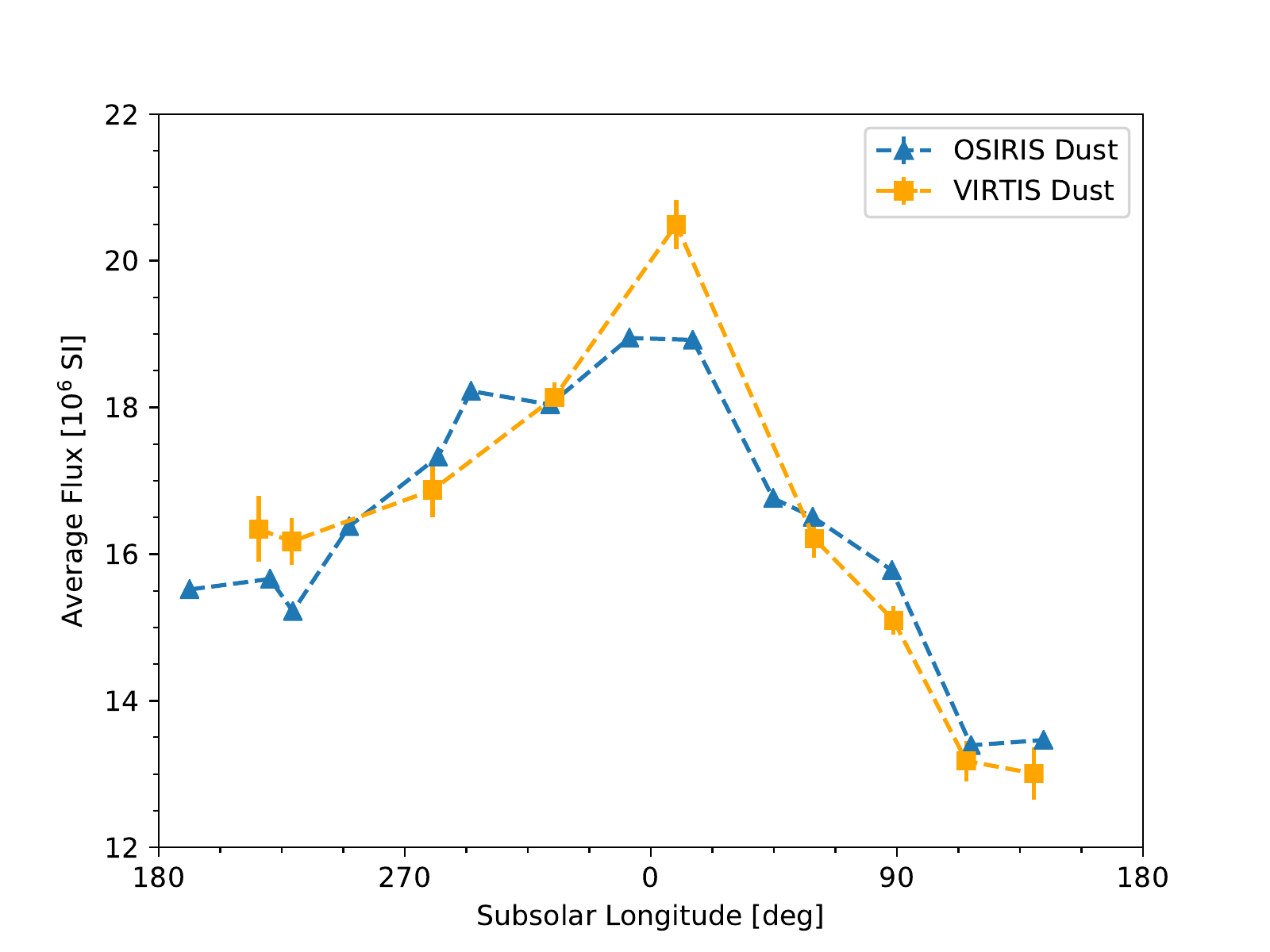}
\caption{Longitudinal variation of the dust flux. Each data point is the average dust flux, inside the $\pm$ 45$\degr$ sector in sub-solar direction at 3.1 km from the comet centre, measured in OSIRIS (blue) and VIRTIS-M (orange) images. Please note the that the error bars associated to the OSIRIS data points are too small to be discerned in the plot.}
\label{fig:osiris_virtis_dust_sector}
\end{figure}

The strongest dust signal is observed in the sub-solar direction (Fig. \ref{fig:images_osiris_virtis}; see also Sec. \ref{sec:azimuthal_results}).
To study the overall dust coma, we have selected a mask with 90$\degr$ angular size ($\pm$ 45$\degr$) in the sub-solar direction, as shown in Fig. \ref{fig:virtis_image_time} (right panel).
We have chosen this angular size to minimize the contribution of fine-scale structures (i.e. jets) present in the images. 

Figure \ref{fig:osiris_virtis_dust_sector} shows the longitudinal variation of the dust flux. Each point of the figure represents the average dust flux inside the $\pm$ 45$\degr$ mask measured in OSIRIS (blue triangles) and VIRTIS-M (orange squares) images. 
The measured average flux in the $\pm$ 45$\degr$ mask is summarised in Table \ref{tab:dust_gas_fraction}.

The OSIRIS and VIRTIS-M measurements are in very good agreement at the wavelength used for the analysis (Fig. \ref{fig:osiris_virtis_dust_sector}).
In addition, the good agreement gives us the possibility to directly compare OSIRIS measurements with VIRTIS-M ones at different wavelengths, without having to consider the possible presence of instrumental effects.
As described in Sec. \ref{sec:dataset_virits}, the analysis of the dust continuum in the VIRTIS-M data is limited by in field stray light when the instrument's slit is partially illuminated by the bright nucleus, as is the case in our observations.
For this reason we used only the OSIRIS images to study the dust in all subsequent sections, avoiding the need to interpolate across stray light regions, as was necessary in previous work \citep{Rinaldi2016}, while using VIRTIS-M to measure the gas. 

\subsection{Dust brightness and $Af\rho$}
\label{sec:dust_prod}

\begin{figure}
\centering
\includegraphics[width=1.0\columnwidth,trim=0.5cm 0cm 0.2cm 1.0cm,clip=true]{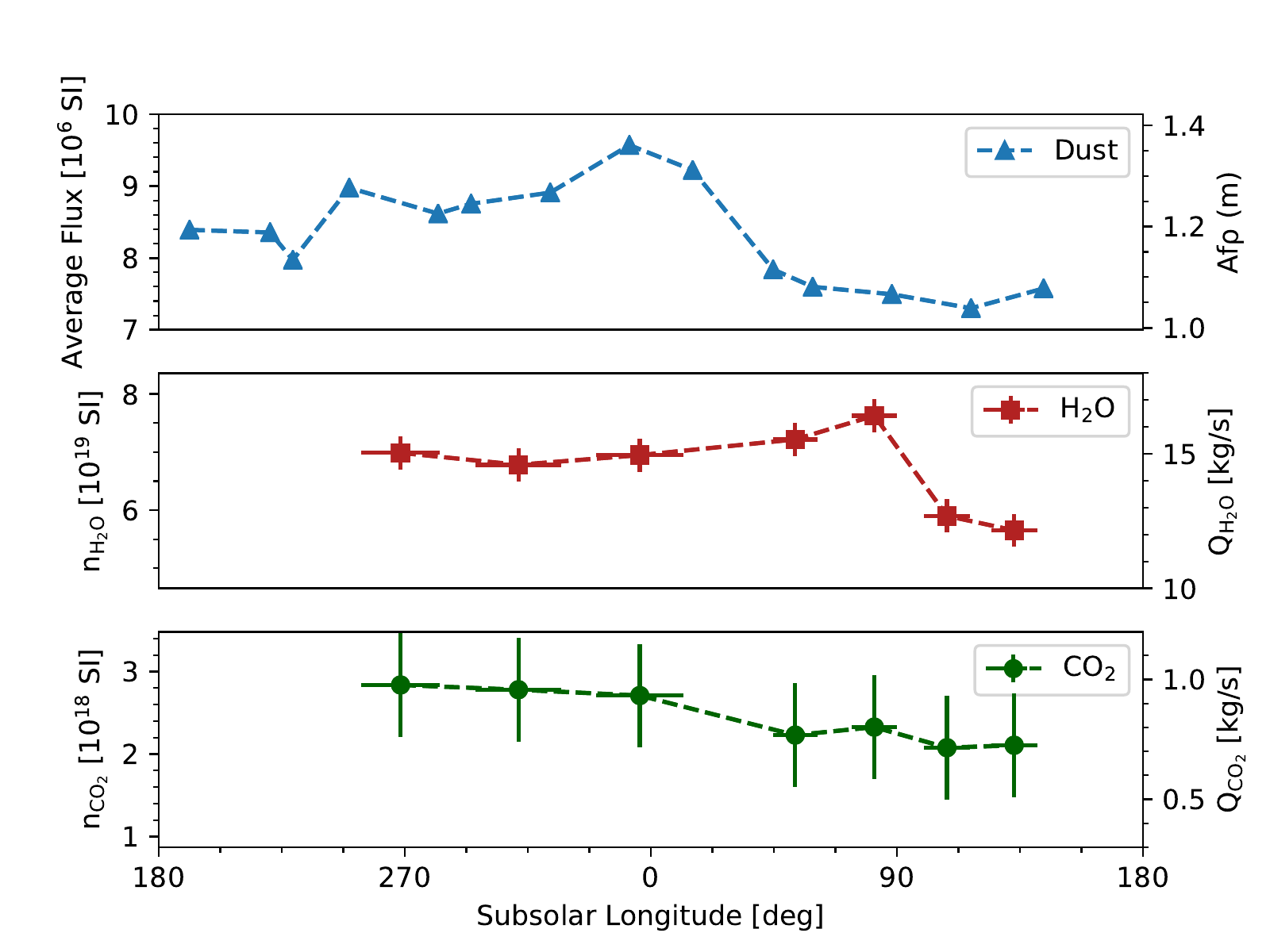}
\caption{Average dust flux and corresponding $Af\rho$ (top panel), H$_2$O (central panel) and CO$_2$ (bottom panel) column densities and corresponding production rates, as functions of sub-solar longitude, in an annulus at 3.1 km from the centre of the comet.}
\label{fig:signal_in_anulus}
\end{figure}

We calculate $Af\rho$, which is commonly used to quantify dust brightness in comets \citep{Ahearn1984} and often used as proxy of dust production; it is proportional to the dust loss rate if the dust size distribution and velocity are constant.
As we do not know the size distribution or velocity we use only the \emph{observed} flux of scattered light to quantify the dust in each image, and include the conversion to $Af\rho$ only to allow convenient comparison with other observations, but do not attempt to derive any absolute dust production rate in kg/s.
The average flux in a full annulus as function of sub-solar longitude is shown in Fig. \ref{fig:signal_in_anulus} and summarised in Table \ref{tab:dust_gas_fraction}. About 50\% of the total dust flux in the full annulus comes from the 90\degr \;sector in sub-solar direction, as shown in Table \ref{tab:dust_gas_fraction}.

To translate the observed scattered light intensity along the Line-of-Sight (LoS) into a {\it local} $Af\rho$ we use the method developed in \citet{FinkRubin2012} and \citet{FinkRinaldi2015}. 

If the considered annulus is at a sufficiently large distance from the nucleus, in the collision-free flowing zone, and no additional production or destruction of dust occurs, the calculated $Af\rho$ can provide a global measure of $Af\rho$ in the immediate vicinity of the nucleus, and it will miss only a small fraction of the total dust emitted. 
\citet{Rinaldi2016} found that closer than 4 km from the surface the dust intensity decreases much faster than $1/\rho$, which implies that the dust acceleration region is sampled by our measurements at 3.1 km from the comet centre. 
\citet{Gerig_2018Icarus} determined that the average starting point of the 1/$\rho$ behaviour is (11.9 $\pm$ 2.8) km.
Since we are not fulfilling the `steady state' condition, the determined $Af\rho$ cannot be directly compared with ground-based global measurements.
Nevertheless, the result we obtain is similar to the ground-based value of $Af\rho \sim 0.9-1.0$ m at this time \citep{Snodgrass2016}. 
We obtained an $Af\rho$ between 1.0 m and 1.4 m, as shown in Fig. \ref{fig:signal_in_anulus}.

\section{Gas production rate: H$_2$O and CO$_2$}
\label{sec:gas_prod}
\subsection{Gas production rate derived from VIRTIS-M data}
\label{sec:gas_prod_VIRTIS}

To determine the gas production rates, we need to calculate the average emitted band intensity inside the 3.1 km annulus and to translate it into a gas column density ($n\:(\rho$)) \citep{Migliorini2016,Fink2016}. 
For these calculations we discard the last image cube (\#9) because of some radiometric problems at the wavelengths close to the H$_2$O gas emission.  
In the image cube \#8 (see Fig. \ref{fig:images_osiris_virtis}) only the dayside part of the coma is observed, so  this image cube is also discarded.
Only the first seven image cubes (\# 1-7) have the full nucleus inside the FOV, allowing to retrieve the complete azimuthal behaviour of the gas in the coma. 

For the combined H$_2$O bands at 2660 and 2730 nm we used the fluorescence efficiency at 1 au $g_0$ = 2.745 $\times$ 10$^{23}$ W/molec. \citep{Bockelee-Morvan2015} and for the CO$_2$ band we use $g_0$ = 1.25 $\times$ 10$^{22}$ W/molec. \citep{Debout2016Icarus}. 
The uncertainty on the H$_2$O and CO$_2$ column density calculations is $\sim$ 10$\%$. 
The measured average water and CO$_2$ column densities as functions of sub-solar longitude are displayed in Fig. \ref{fig:signal_in_anulus} (centre and bottom panels) and summarised in Table \ref{tab:dust_gas_fraction}.
At 3.1 km from the comet centre 
78\% of the water column density is contained within an angle of $\pm$90\degr \;and 52\% within an angle of $\pm$45\degr \;in the sub-solar direction. 
There is very little scatter in the percentages in the first seven observations despite the different configurations of nucleus, Sun and spacecraft.
The CO$_2$ distribution does not follow the direct solar illumination and has essentially no correlation with the water column density distribution (Fig. \ref{fig:signal_in_anulus}). 
The CO$_2$ molecules are emitted mostly from the Southern hemisphere of the comet as shown in Fig. \ref{fig:images_osiris_virtis} (fourth column). 
This is the reason why 50\% of the CO$_2$ column density is contained within an angle of $\pm$90\degr \;in the sub-solar direction (dayside) and the same percentage is on the nightside, with little variability in those percentages (Table \ref{tab:dust_gas_fraction}).
The large variability of the CO$_2$ column density, from 19\% to 38\%, in the $\pm$45\degr sector in sub-solar direction is due to  the orientation of the comet’s spin axis during these observations.

To convert the gas column densities into production rates we use the method described in \citet{Fink2016}. 

The determined gas production rates are listed in Table \ref{tab:prod_rate}.
\citet{Fink2016}, for the same dataset, analysed the emission intensity of CO$_ 2$ and H$_ 2$O and their distribution in the coma using a slightly different annulus at 2.8 km from the centre to the comet. Our result obtained here for the H$_2$O and $\mathrm{CO}_2$ distribution are in good agreement with the findings of \citet{Fink2016} and the gas spatial distribution maps obtained by \citet{Migliorini2016}.

It is evident from Fig. \ref{fig:signal_in_anulus} that there is little correlation between the measured productions of dust, water, and $\mathrm{CO}_2$. 
In addition, none of the patterns correspond to the variation of the cross section area of the illuminated nucleus (Fig. \ref{fig:thermal_model} bottom panel).
The lack of correlation between dust and $\mathrm{CO}_2$ is not unexpected, indicating the indistinct role of $\mathrm{CO}_2$ outgassing in driving the global dust emission.
On the other hand, the discrepancy between the patterns of dust and water measurements is less intuitive. Water outgassing is the dominant driver of dust activity from the northern hemisphere \citep{DeSanctis2015,Shi2018}, prior to the observations analysed here (days before Northern autumn equinox).
The deviation of water production from the variation of illuminated cross section, in particular, suggests that the nucleus surface is not homogeneously active as presented by \citet{Marschall2019Icarus}.
In the following section, we perform a simple thermo-physical analysis in order to shed some light on this discrepancy.

\subsection{H$_2$O production rate computed by a thermo-physical model}
\label{sec:thermo_model}
\begin{figure}
\centering
\includegraphics[width=\columnwidth,trim=2cm 6cm 0cm 0.0cm,clip=true]{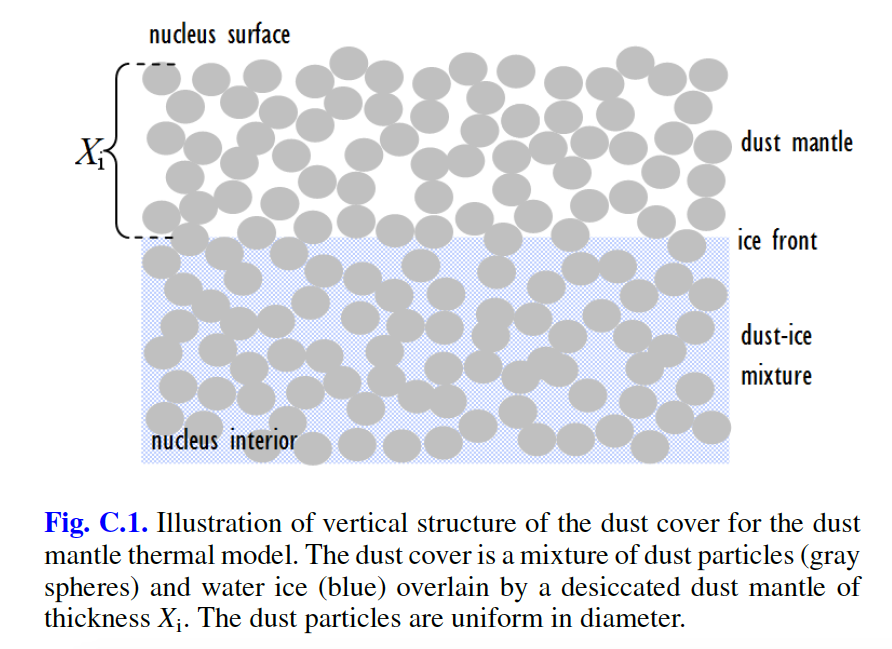}
\caption{Illustration of the dust cover assumed in the thermo-physical model. Figure adapted from \citet{Hu2017AA}. $X_i$ is the thickness of the desiccated dust mantel.}
\label{fig:sketch_model}
\end{figure}

We wish to compare the observed pattern of dust and gas release to a simplified model that describes what would be expected in the case of homogenous activity controlled only by illumination of each surface element. In order to do this 
we employed a thermo-physical model \citep{2017MNRAS.469S.295H} to estimate the total water production rate of the nucleus over a full rotation of 67P that encompasses all observations described above.  

\subsubsection{Model description}

The shape of the nucleus is approximated by a  model consisting of 1500 facets \citep{2015AandA...583A..33P}. We assumed the nucleus is covered by a desiccated dust layer, or dust mantle, of constant  thickness and composed of uniform spherical dust aggregates. Water ice is present underneath the dust mantle (Fig. \ref{fig:sketch_model}). 
The sublimation flux is strongly influenced by the temperature of the ice front. The temperatures of the nucleus subsurface as a function of depth are estimated via the solution of the 1-D heat equation, balancing the input energy from solar illumination of the surface with heat re-radiated, conducted into the surface or used in sublimating ice.

However, the nucleus interior (beneath the mantle) is not composed of pure water ice and, thus, sublimation cannot take place everywhere. For this reason, the factor $f_\mathrm{i} \in (0,1)$ is introduced that measures formally the areal fraction of water ice and is approximately inverse to the dust-to-ice ratio of the subsurface \citep{1997Icar..130..549C}.
We assumed that heat flux vanishes beyond several (diurnal) skin depths.

At any given epoch, the position vector of the Sun with respect to (the body-fixed frame of) the nucleus is obtained via the SPICE kernels\footnote{The SPICE kernels are available from ESA at https://www.cosmos.esa.int/web/spice/spice-for-rosetta} for 67P. It is subsequently transformed into local horizontal coordinates via a series of rotations of the coordinate system to yield the solar incidence angle. We made use of a ``Landscape'' database for 67P that delineates the skyline at each location on the nucleus (each facet of the shape model) in order to determine efficiently the local illumination \citep{2017MNRAS.469S.295H}.

The 1-D heat equation is solved via the Crank-Nicolson method. The solutions are diurnally equilibrated temperatures $T$ and water production rates $Z$ that repeat or coincide, if the seasonal cycle is neglected, with those exactly one comet rotation apart, e.g., $T(t \pm t_\mathrm{p}) = T(t)$ and $Z(t \pm t_\mathrm{p}) = Z(t)$ with $t_\mathrm{P}$ being the rotation period of 67P.

\subsubsection{Choice of model parameters}

\begin{table}[t]
\caption{Parameters for thermo-physical modeling.}
\label{table:thermparam}
\footnotesize\centering
\begin{tabular}{lcrl}
\hline
\hline
\\
Parameter & Symbol & Value \\
\\
\hline
\hline
\\
Bond albedo & $A_\mathrm{B}$ & 0.01 \\
Emissivity & $\varepsilon$ & 1 \\
Heat conductivity [$\text{W}\,\text{m}^{-1}\text{K}^{-1}$] & $\kappa$ & $2 \times 10^{-3}$ \\
Specific heat capacity [$\text{J}\,\text{kg}^{-1}\text{K}^{-1}$] & $c$ & 1000 \\
Density [$\text{kg}\,\text{m}^{-3}$] & $\varrho$ & 500\\
Diameter of dust aggregate [mm] & $d_\mathrm{P}$ & 1 \\
Thickness of dust mantle [mm] & $X_\mathrm{i}$ & 5\\
Area fraction of ice & $f_\mathrm{i}$ & 0.01 & \\
\hline
\end{tabular}
\end{table}

A summary of the key parameters of the thermo-physical model is given in Table \ref{table:thermparam}.
For simplicity, it is assumed that the nucleus subsurface is homogeneous. We also neglect variability of the parameters, e.g., changes of mantle thickness and loss of ice underneath. With such assumptions, the model parameters are treated as constants.
Following the argument by \citet{2017MNRAS.469S.755B} that the dust aggregates, though clearly non-uniform in size, are of the order of a few millimetre in diameter, we adopt the diameter of the dust aggregates $d_\mathrm{P} = 1$~mm. 
The OSIRIS observation of the dust activity that continued for about one hour after sunset indicates that water ice was present at some depth of less than 1~cm \citep{2016A&A...586A...7S}. 
The long-term (seasonal) evolution of the total water production of 67P throughout perihelion can also be modelled with a mantle thickness of 5~$\leq X_\mathrm{i} \leq$ 10~mm. 
Overall, water ice is rarely exposed \citep{2015Sci...347a0628C}. 
When detected, it is usually present in small quantities, e.g., a few percent \citep{2015Natur.525..500D, 2016Natur.529..368F, 2016A&A...595A.102B}. 
There is observational evidence that the average water ice abundance in the top $\sim1$~m over the northern hemisphere can not exceed 10\% \citep{Hu2017AA}. 
Therefore, the area fraction of water ice ($f_\mathrm{i}$) is probably similar.
The scarcity of ice suggests that the thermo-physical properties of the nucleus are dominated by those of the refractory component. 
We are aware that the thermal parameters, namely, the conductivity ($\kappa$) and specific heat capacity (c), are dependent on temperature. 
In particular, thermal radiation through the pores enhances the efficiency of heat transfer. 
In such a case, the radiative component of the conductivity varies with $T^3$. 
However, the radiation affects most significantly temperatures below the diurnal skin.
In the case of millimetre-sized particles and assuming ice sublimation mostly from above the diurnal skin depth, the enhancement of water production by radiation is not notable \citep{Hu2019}. 
In addition, the dependence of heat capacity on temperature is linear \citep{Orosei1995}, while the exact behaviour of material on 67P is largely unknown.
Hence, we neglect the temperature dependence of the parameters and adopt $\kappa_\mathrm{d} = \kappa_\mathrm{i} = 0.002\ \mathrm{W\,m^{-1}K^{-1}}$ (conductivity of the dry dust mantle and the underlying icy dust, respectively), 
$c=1000\,\mathrm{J\,kg^{-1}K^{-1}}$ and $\varrho=500\,\mathrm{kg\,m^{-3}}$, which corresponds to a thermal inertia of 30~$\mathrm{W\,m^{-2}K^{-1}s^{1/2}}$, as measured by MIRO \citep{Schloerb2015}.

\subsubsection{Model results}

\begin{figure}
\centering
\includegraphics[width=1.0\columnwidth,trim=0.5cm 1.3cm 0cm 1.0cm,clip=true]{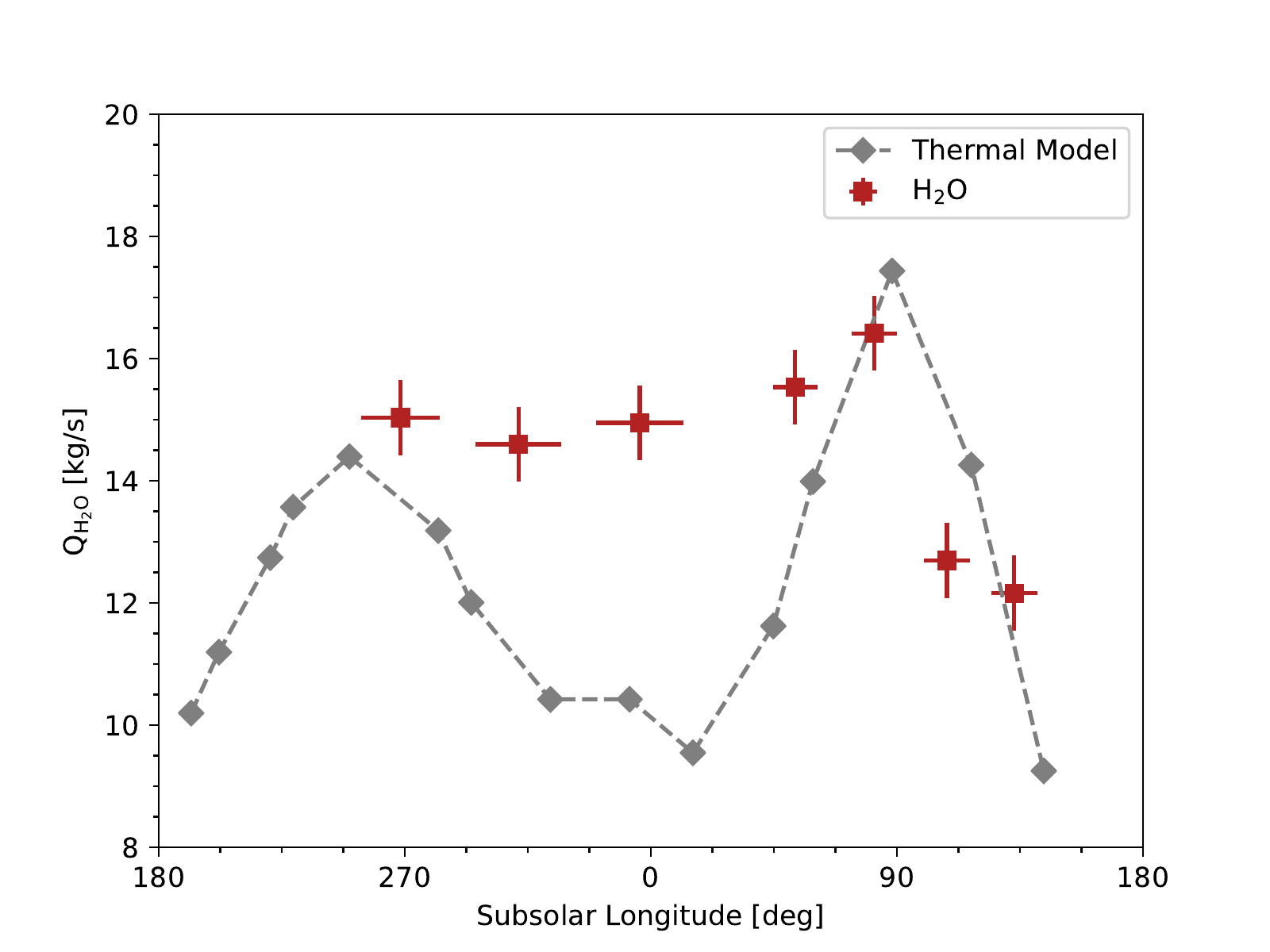}
\includegraphics[width=1.0\columnwidth,trim=0.5cm 6.5cm 0cm 1.0cm,clip=true]{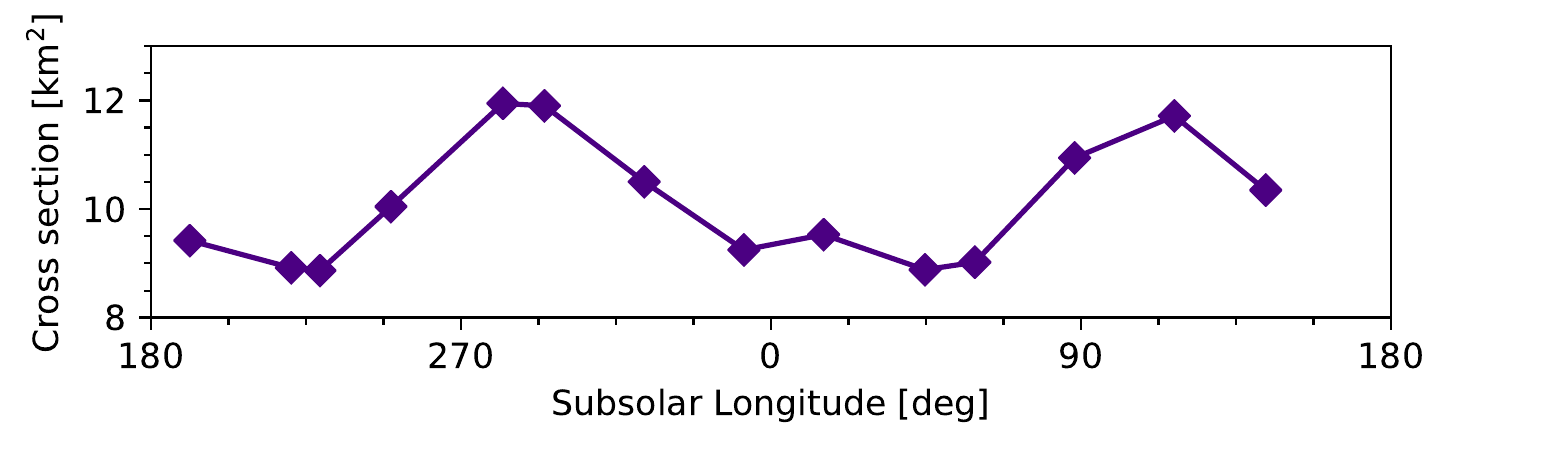}
\caption{Top panel: Comparison between the water production rate as output of the thermo-physical model and the one measured in VIRTIS-M-IR spectra. Bottom panel: Illuminated nucleus cross-section at the time of each OSIRIS observation.}
\label{fig:thermal_model}
\end{figure}

The water production rates at each observation time, from both the output of the thermal model and the VIRTIS-M observations described in Sec. \ref{sec:gas_prod_VIRTIS}, are shown in Fig. \ref{fig:thermal_model}.
Regulated by the mantle thickness and ice abundance, the modelled water production rate is in good agreement with the measurements in terms of overall magnitude.
The variation largely follows that of the cross-section area of the illuminated nucleus, as expected for a homogeneous nucleus activity model. A phase shift of about 20\degr\ in sub-solar longitude is evidently attributable to the presence of the dust mantle causing a thermal lag of about half an hour at the depth of the ice front (i.e., 5~mm).
However, there is a clear underestimation in the modelled production rate compared with the measurements in the sub-solar longitude range (270\degr -- 50\degr).
This indicates that the real gas production varies between areas and must depend on local (sub)-surface properties, not just illumination and topography of the nucleus; this is discussed further below.

\section{Azimuthal gas and dust distribution}
\label{sec:azimuthal_results}

\begin{figure}
\centering
\includegraphics[width=0.49\columnwidth]{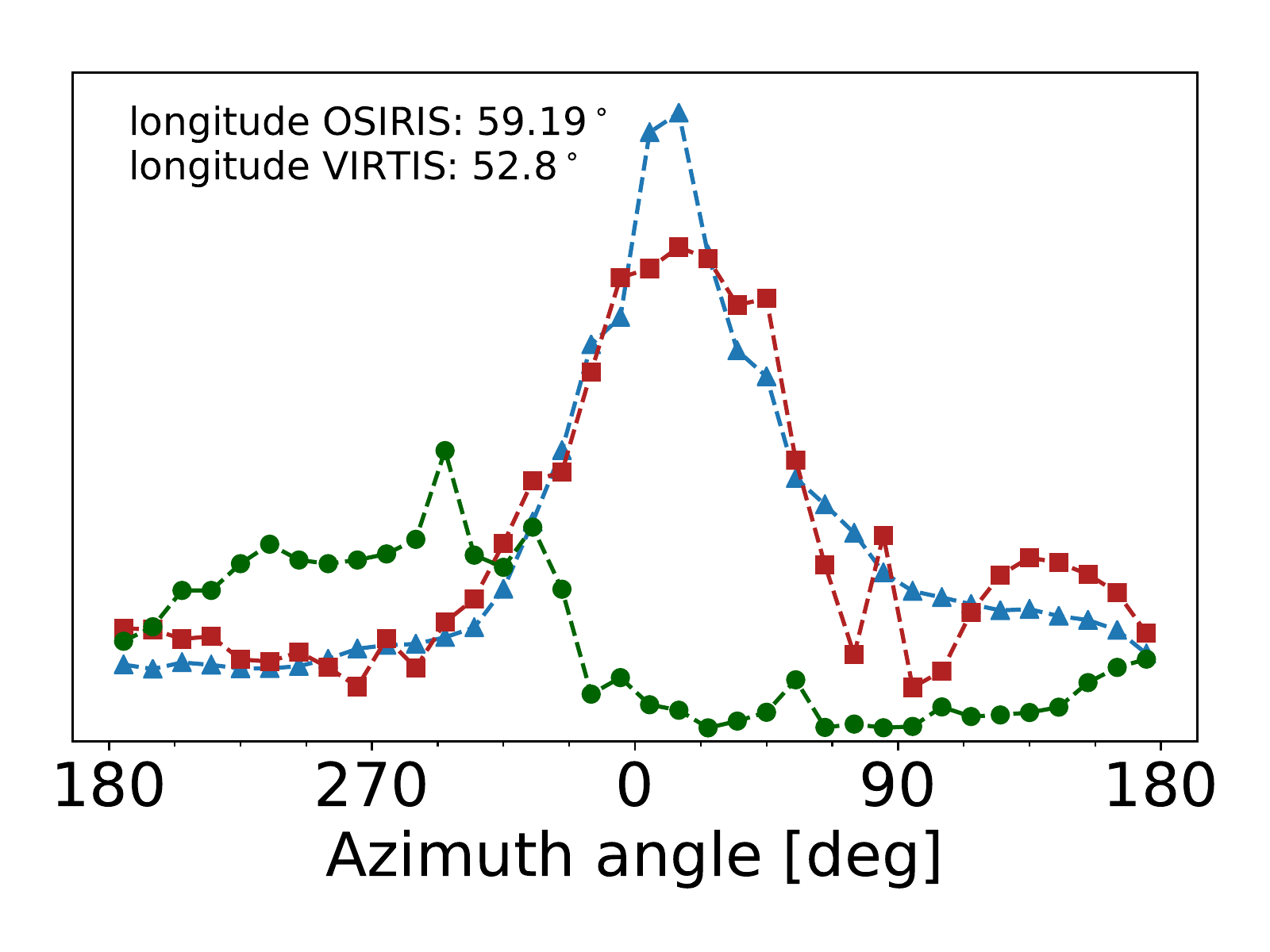}
\includegraphics[width=0.49\columnwidth]{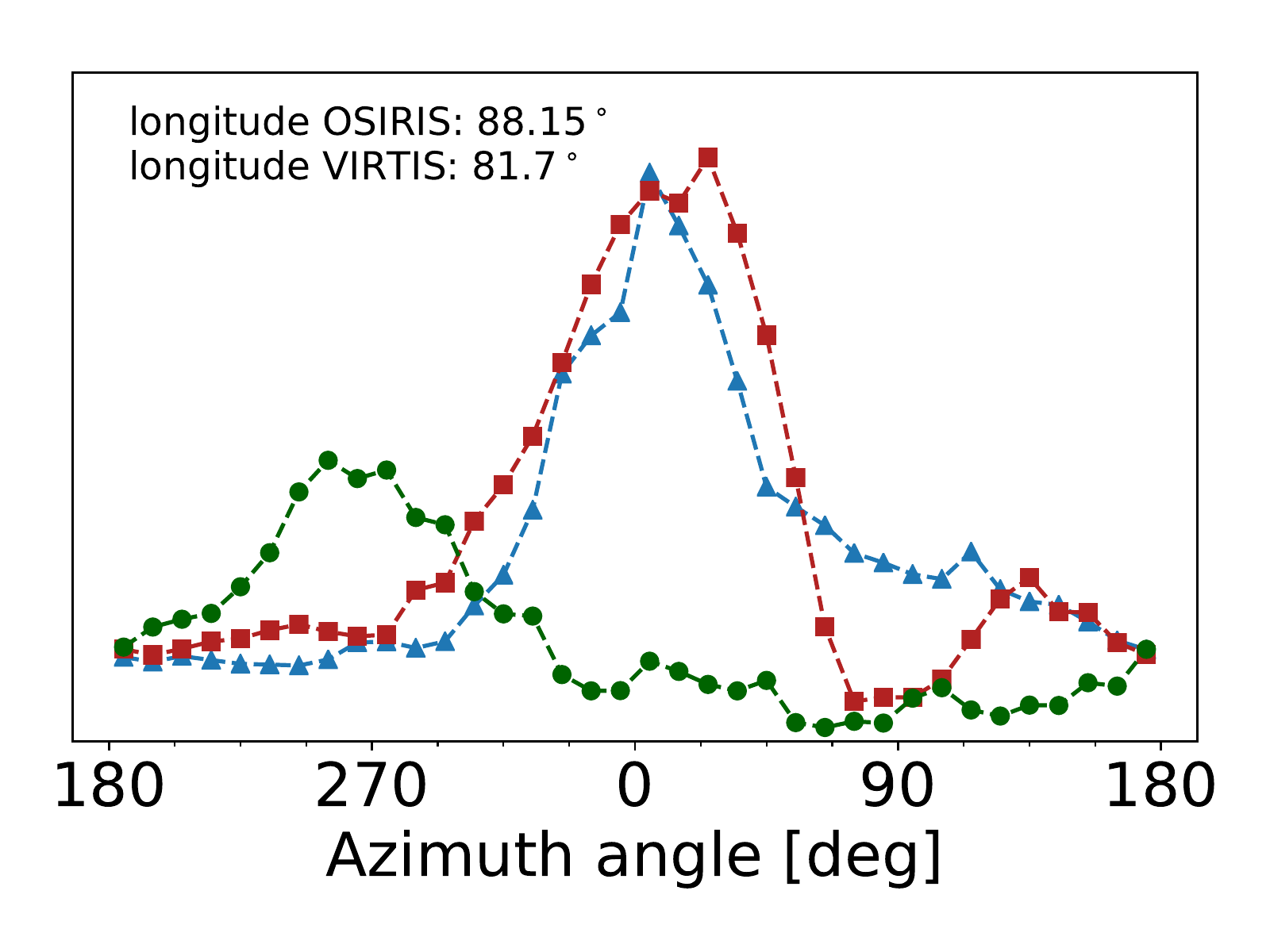}
\includegraphics[width=0.49\columnwidth]{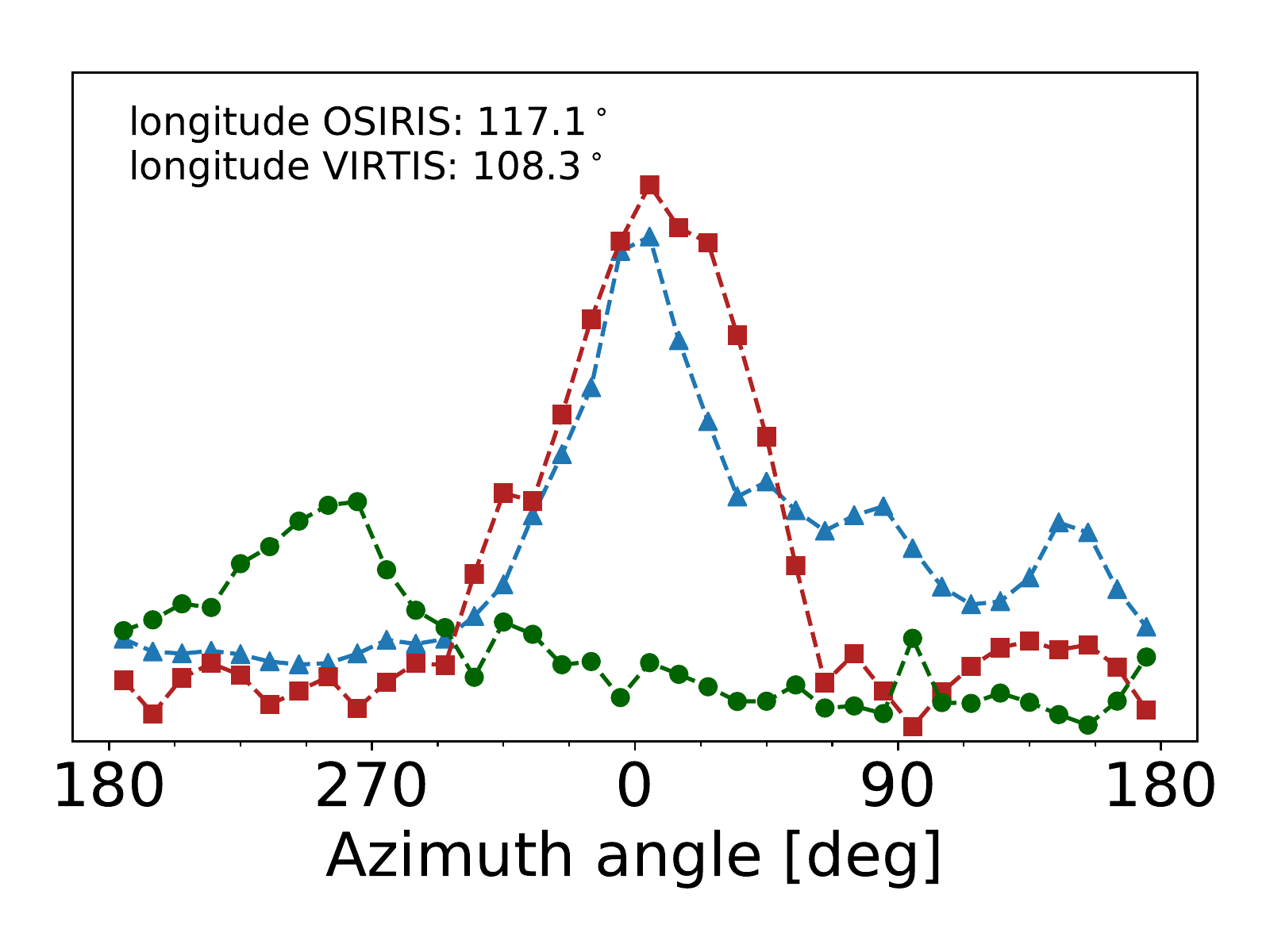}
\includegraphics[width=0.49\columnwidth]{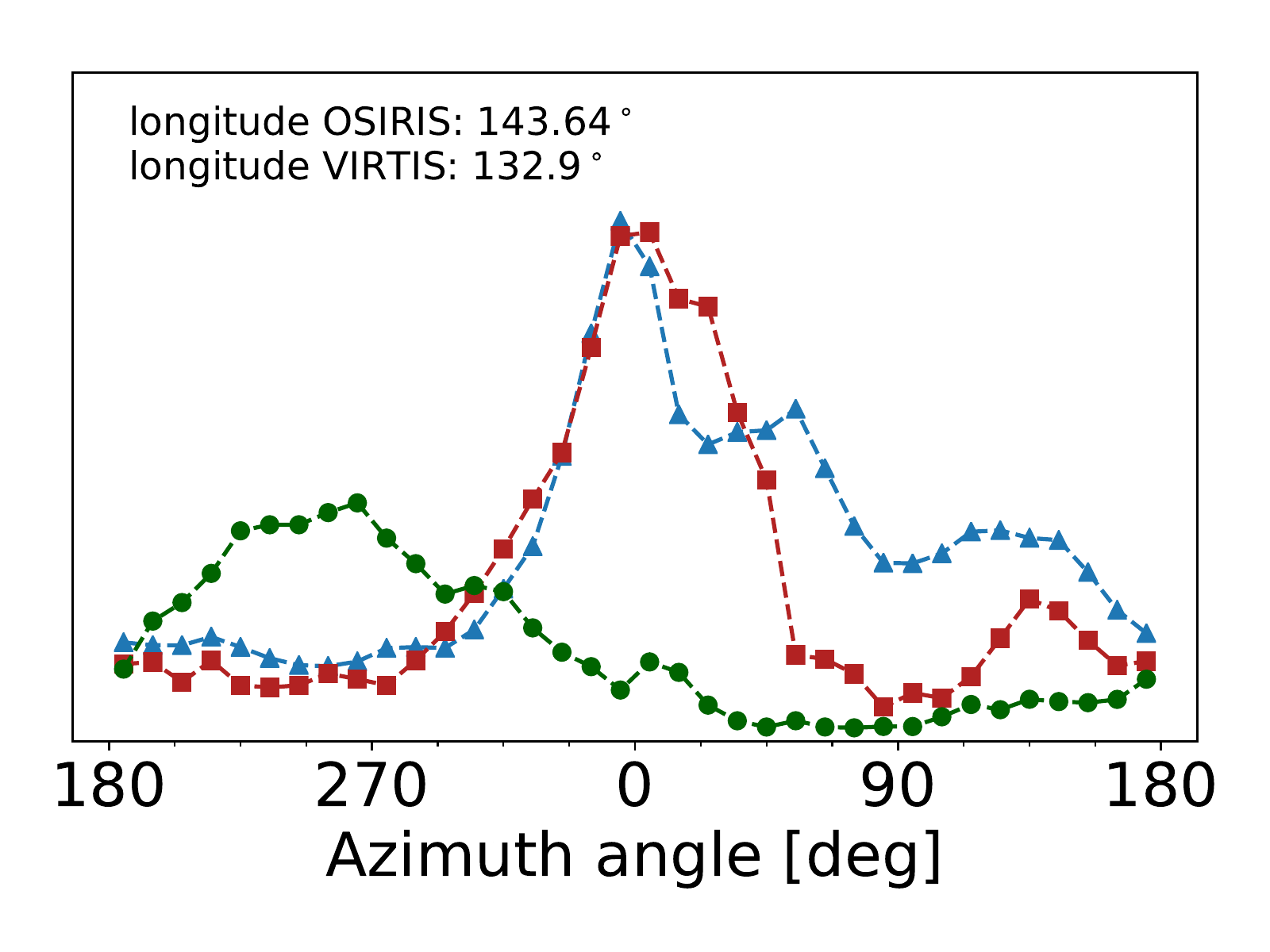}
\includegraphics[width=0.49\columnwidth]{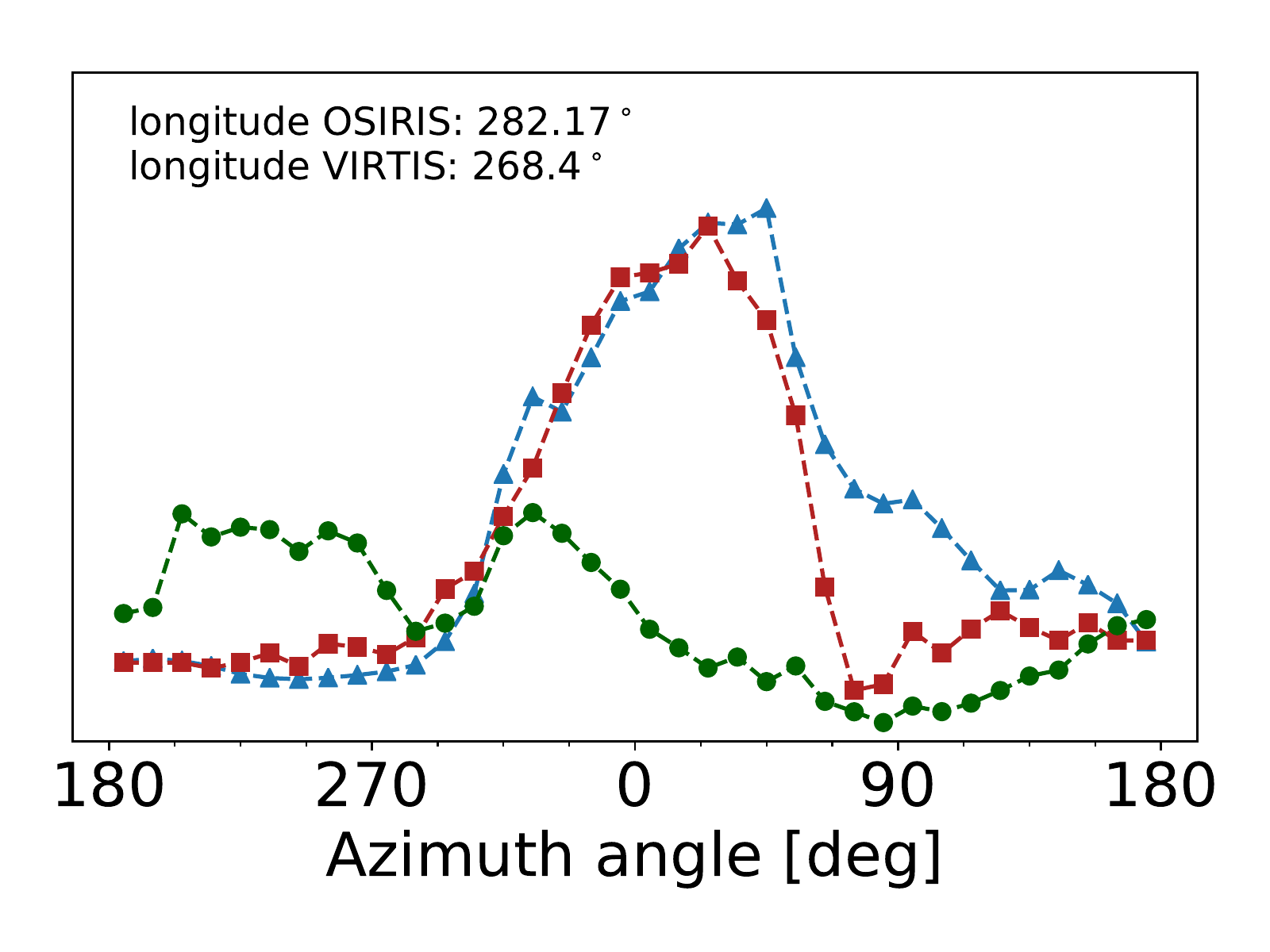}
\includegraphics[width=0.49\columnwidth]{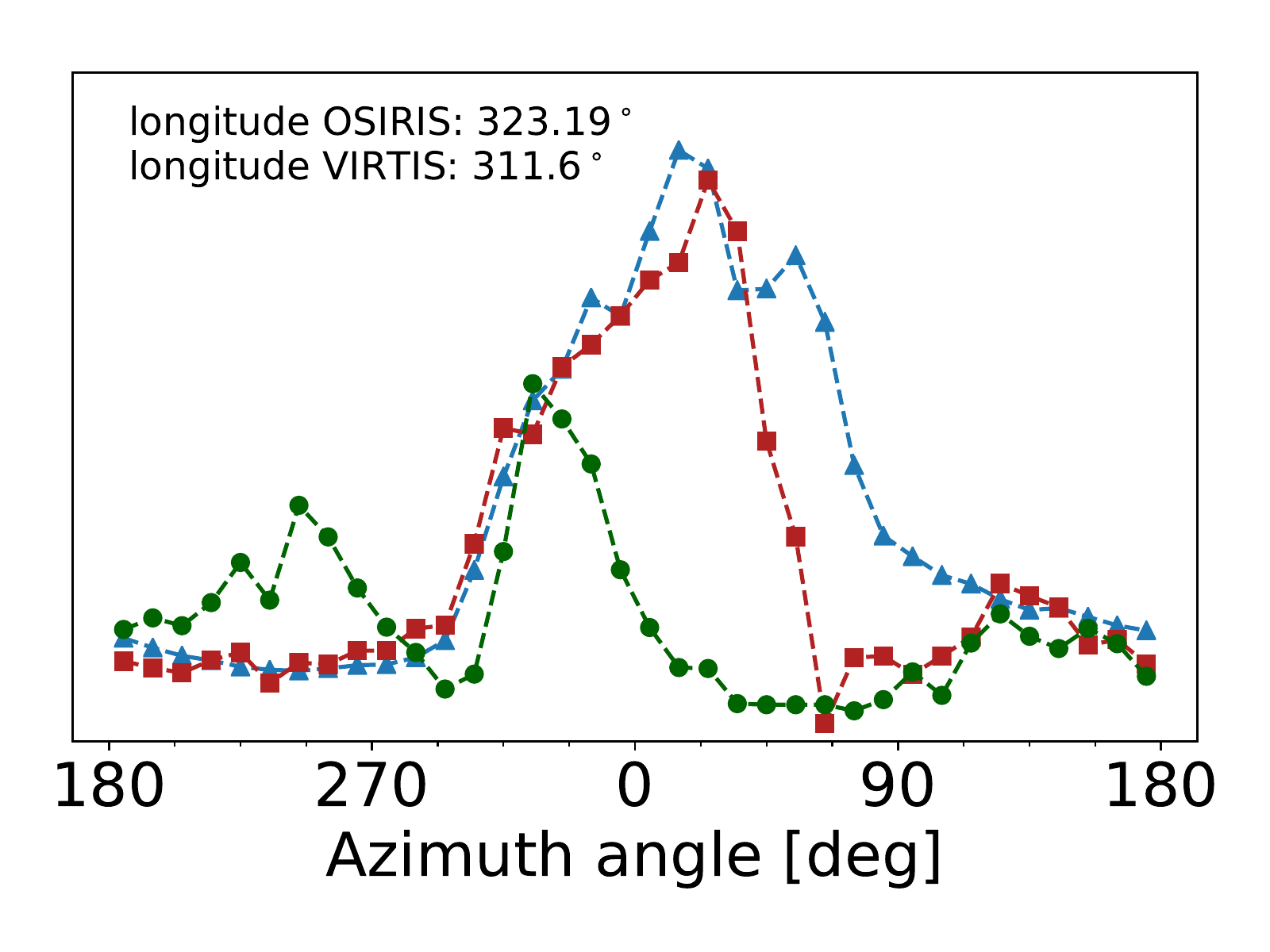}
\includegraphics[width=0.49\columnwidth]{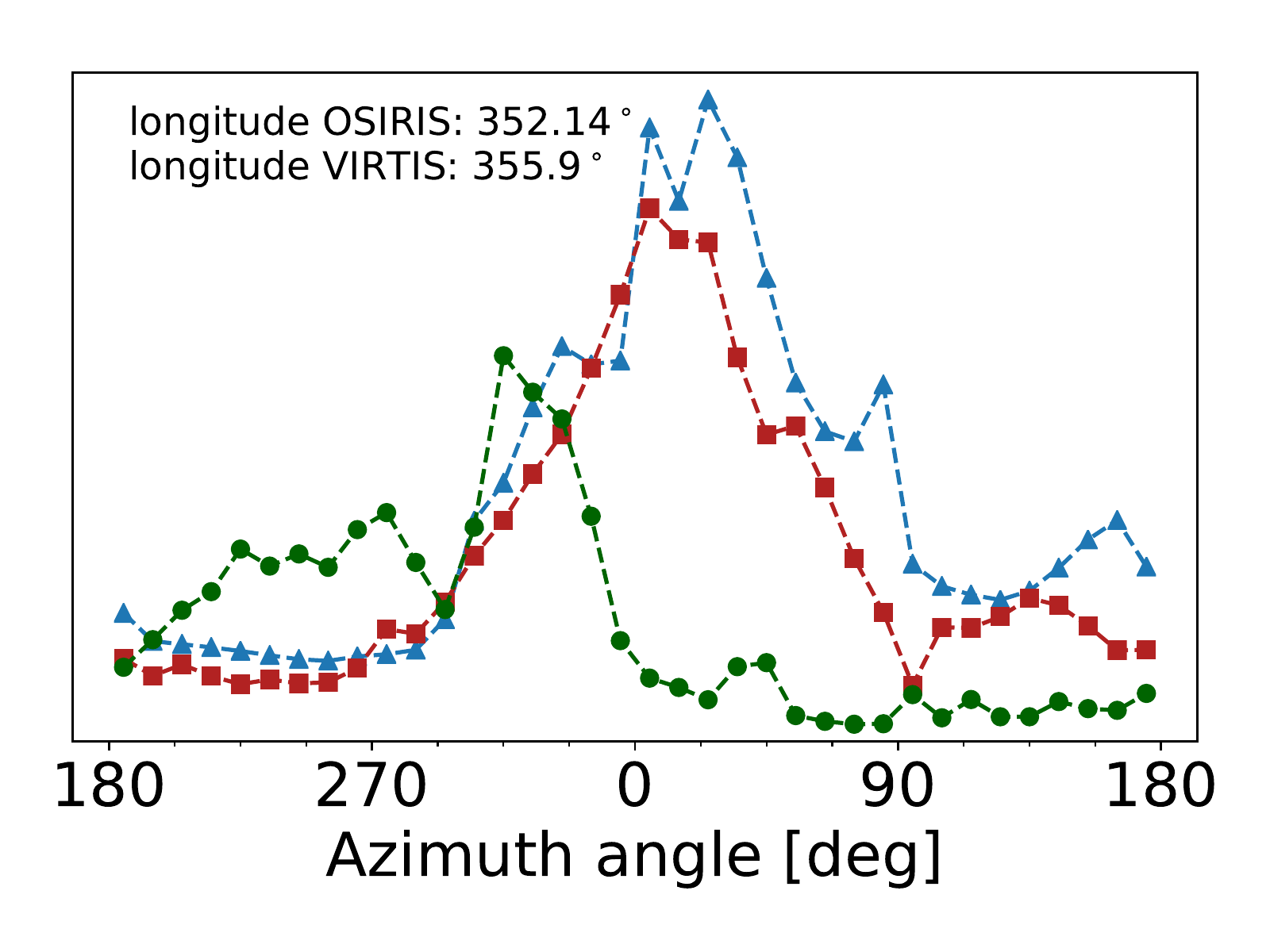}
\caption{Azimuthal profiles of dust brightness and gas band intensity. 
OSIRIS dust is shown as blue triangles, VIRTIS-M $\mathrm{H_2O}$ as red squares, and VIRTIS-M $\mathrm{CO_2}$ as green circles. 
For displaying purposes, VIRTIS-M $\mathrm{H_2O}$ and $\mathrm{CO_2}$ band area intensities are scaled to the OSIRIS dust values. }
\label{fig:azimuthal_profiles}
\end{figure}

Figure \ref{fig:azimuthal_profiles} shows the angular distribution of water (red squares) and CO$_2$ (green circles) band intensity, measured in the first 7 VIRTIS-M image cubes, and the angular distribution of the dust brightness (blue triangles), measured in the OSIRIS images closest in the sub-solar longitude to the VIRTIS-M data.
In all observations, the absolute maximum for dust and water vapour is located in the sub-solar direction. 
A secondary water vapour peak occurs at roughly 130\degr\ and it is related to the variable illumination of the neck area. 
CO$_2$ peaks between 180\degr\ and 270\degr, consistent with where the Southern hemisphere is located (Fig. \ref{fig:images_osiris_virtis}).

The spatial (or angular) correlation between dust and water, both coming from the sub-solar side of the comet, and already observed by \citet{Rinaldi2016}, shows that water is the main driver of dust activity in this time period.
This is also generally consistent with observations from the ground, showing that long-term variations in total water production rate correlate with the total dust brightness \citep{Hansen2016}.

The presence of CO$_2$ ice in the these regions of the Southern hemisphere and in the same period is not unexpected, since VIRTIS detected a CO$_2$-ice rich area in the Anhur region at the end of March 2015 \citep{Filacchione2016Science}.
On the Southern hemisphere, the CO$_2$ ice is generally closer to the surface, thus more accessible.
When the comet comes close to the Sun, the Southern summer is very intense. 
In this strong sunlight, erosion rates are fast enough to expose fresh, primordial layers of the interior, which are rich in CO$_2$.
Quantitative calculations of the comet erosion rates during its orbit are presented by \citet{Keller2015}.
The increase of CO$_2$ abundance in the coma during the perihelion time-frame is analysed in \citet{Bockelee2016}. 
The authors found that during the aforementioned period the measured abundance ratios of CO$_2$ increased by a factor of 30 with respect to what was found above the illuminated Northern hemisphere.

\section{Discussion}
\label{sec:discussion}

\begin{figure}
\centering
\includegraphics[width=1.0\columnwidth,trim=0.0cm 0cm 0.0cm 0.0cm,clip=true]{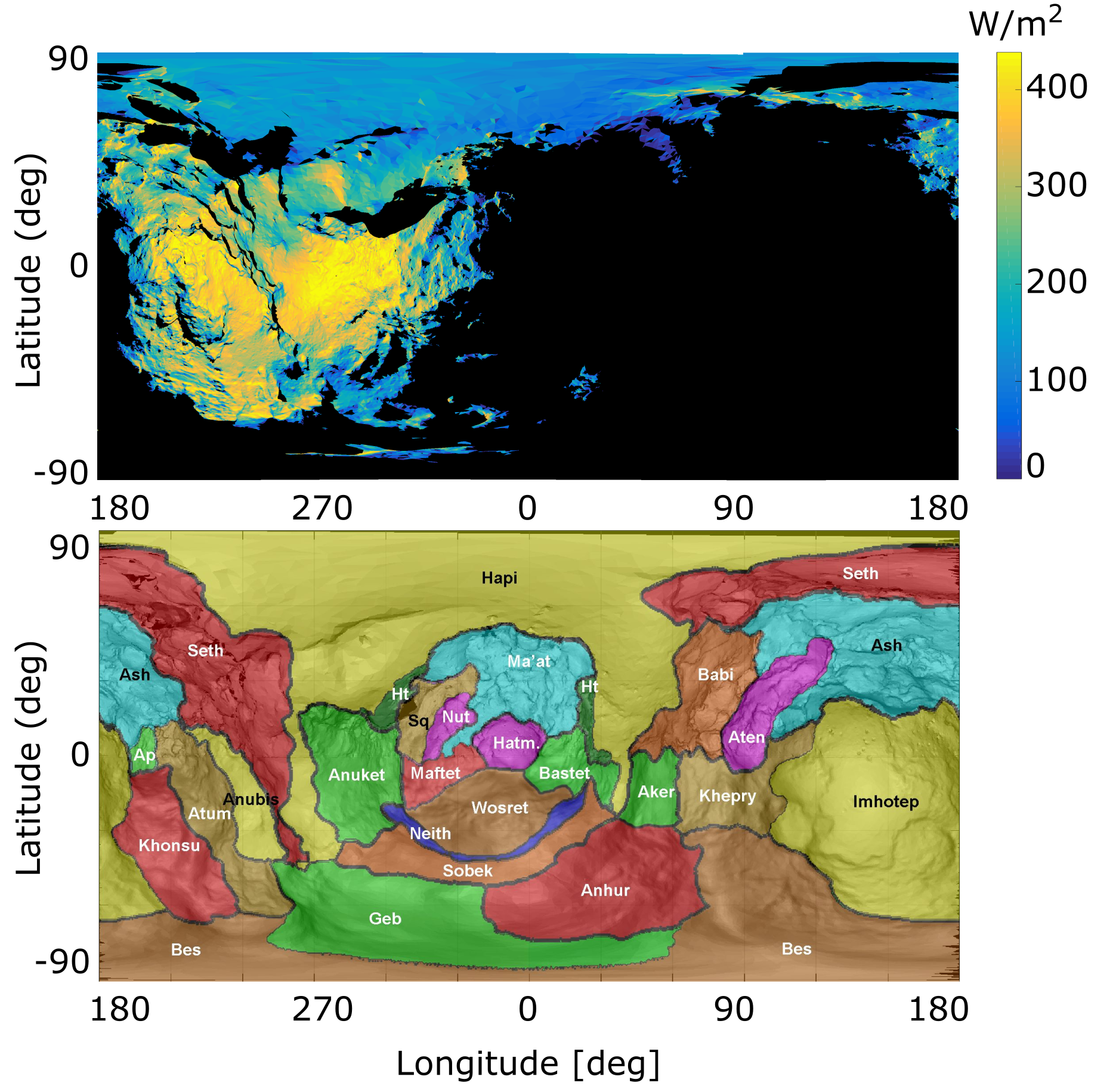}
\caption{Top panel: Example of insolation map, calculated for 09:25:32 UTC. Bottom panel: Region map of 67P adapted from \citet{El-Maarry2016AA}.}
\label{fig:illumination_map_example}
\end{figure}

\begin{figure*}
\centering
\includegraphics[width=1.0\textwidth,trim=0.1cm 0.5cm 0.0cm 0.0cm,clip=true]{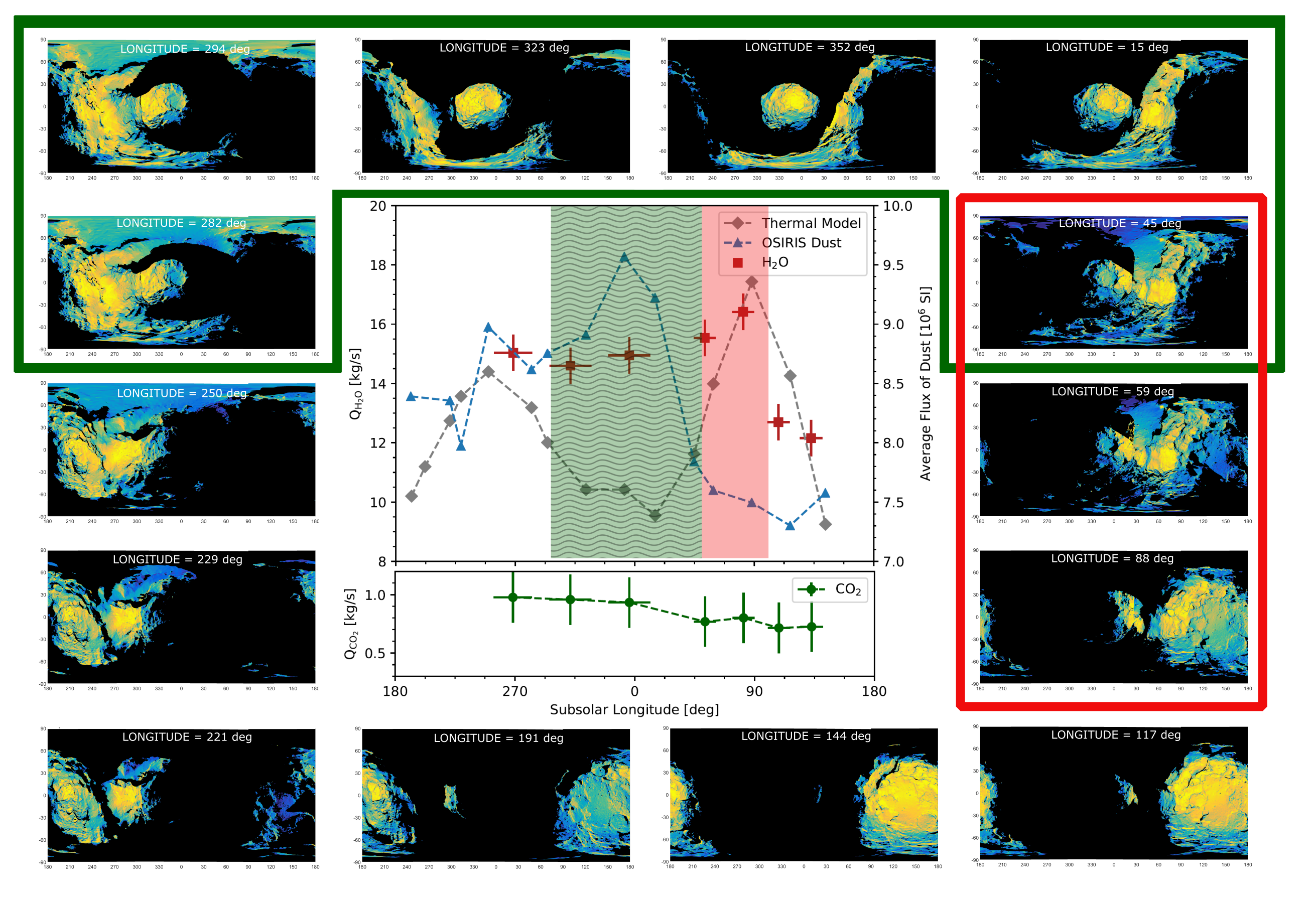}
\caption{Insolation maps for the time of each OSIRIS observation. Inset: total dust brightness and gas production rates, measured in the full 3.1 km annulus, as functions of sub-solar longitude. The green box encloses the insolation maps corresponding to the sub-solar longitudes where an excess of measured water production was found, relative to a simple model. The red box frames the insolation maps corresponding to the sub-solar longitudes where a minimum of dust and a maximum of water production were measured.}
\label{fig:illumination_map}
\end{figure*}

To understand the illumination conditions across the surface of 67P's nucleus at the time of each OSIRIS observation, we generated insolation maps.
One example is shown in Fig. \ref{fig:illumination_map_example} (top panel), along with a map identifying the different regions of the nucleus (bottom panel).
The distribution of solar irradiance is calculated on a polyhedral shape model representing the nucleus of 67P with 499,902 facets \citep{2017A&A...607L...1P}. 
For each epoch, the Sun's position in the Cheops body-fixed frame of the comet \citep[as defined in][]{2015AandA...583A..33P} is derived from the reconstructed ephemeris and rotational status of 67P using the SPICE tool-kit \citep{Acton1996}. 
Note that the concave shape of 67P prevents the display of certain surface areas in maps with equidistant cylindrical projection. 
However, the overall pattern of illumination shown in the figure is not affected.

Figure \ref{fig:illumination_map} displays the insolation maps at the time of each OSIRIS observation.
In the central inset we show the modelled and observed H$_2$O production rates (grey and red curves, respectively), and the observed CO$_2$ production rate (green curve), and the total dust brightness (blue curve), calculated inside the full 360\degr\ annulus (see Table \ref{tab:prod_rate}).
There is no strong temporal correlation between total dust brightness and water production rates (neither observed nor from the simple model).
It should however be emphasized (as noted in Sec. \ref{sec:azimuthal_results}), that water is still the main driver of dust activity in this time period.
Only the ratio between water production and dust activity is changing.

The green box in Fig. \ref{fig:illumination_map} highlights the excess of water production (red curve) compared to the simple homogeneous model (grey curve) at sub-solar longitudes between 270\degr\ and 50\degr.
The green lines emphasize the corresponding insolation maps.
This corresponds to epochs when the head lobe and regions of the Southern hemisphere with strong seasonal variations (e.g. Bes, Geb, Anhur) are illuminated.
While we caution against over-interpreting the differences between our simple thermal model and the observed water production, it could be argued that these are most easily explained by a higher activity of these Southern regions with respect to the North.

The observed dust brightness (blue curve) shows a pronounced maximum around 0\degr\ sub-solar longitude, which is not pronounced in the water production (red curve).
Previous studies have already shown that the Anhur and Bes regions -- illuminated at this time -- are highly active and sources of several jets \citep{OSIRIS_MPS_2016_Vincent_MNRAS, 2017MNRAS.469S..93F}, thus in agreement with our findings.

At the same epoch (green region) there is also a maximum of CO$_2$ production.
The analysis of the azimuthal profiles in Sec. \ref{sec:azimuthal_results} (Fig. \ref{fig:azimuthal_profiles}) show that the dust is correlated with water and not CO$_2$.
The increased CO$_2$ production in this epoch is therefore not responsible for the peak in the dust activity.

The largest discrepancy between dust and gas production rates can be observed in the red box of Fig. \ref{fig:illumination_map}.
The dust brightness drastically decreases in this sub-solar longitude range (50\degr -- 90\degr), while instead the water production (measured and from the model) displays a maximum. 
These epochs correspond to when Northern consolidated regions \citep{ThomasEtAl2018} (e.g. Bastet, Aker, Khepry, Aten, Babi -- see Fig. \ref{fig:illumination_map_example}, bottom panel) are illuminated and the Southern hemisphere regions with strong seasons are in shadow.

This temporal non-correlation (red box) can either be explained by regional variations of \emph{surface properties} or regional variations of the lifted dust particles' \emph{scattering properties}.
More specifically, we will discuss here the effect of regional variations of \emph{(a)} thickness of the desiccated layer, \emph{(b)} surface cohesion, and the presence of large particles affecting \emph{(c)} gas coupling, thus lifting, and \emph{(d)} permeability of the dust layer.
For the dust particles' scattering properties we will discuss \emph{(e)} lifted dust particle composition and \emph{(f)} lifted dust particle size.

\emph{(a)} We do not attribute the discrepancy of dust and gas production in the red box to the thickness of the desiccated layer on the nucleus surface, as a thicker layer would also quench gas production.
\emph{(b)} A higher cohesion of surface material could quench dust activity \citep{Bischoff2019}, where the water vapour would escape without lifting dust.
This is possible and likely in consolidated regions, which consist of cm-sized pebbles.
\emph{(c)} Even if large (decimetre to metre-sized) particles were easily lifted against their cohesion, they still carry a high mass inertia.
Their low size-to-mass ratio could prevent large enough particles from being carried into the coma from the gas drag.
Fallback for particles in the considered size range was observed by \citet{AgarwalEtal:2016} and many more of these might not even be considerably lifted.
\emph{(d)} If the upper dust layer is dominated by large particles, the gas permeability would also be enhanced.
For granular materials with macroscopic voids, \citet{GundlachEtAl2011} have shown that the gas permeability increases with the size of the constituent particles.
A layer of decimetre-sized dust particles \citep[cp.][]{Pajola2017} would thereafter have a ten times higher permeability than a layer of the same thickness of centimetre-sized particles.
A higher gas permeability of the dust layer would result in a reduced pressure build-up and thus reduced dust production.
The gas could simply escape through the large voids.
Decimetre to metre-sized particles, present in fallback regions (e.g. Ma'at), are almost entirely cleaned-up in consolidated regions, exposing the underlying consolidated material.
This is consistent with the idea of ``self-cleaning'' of the Northern hemisphere proposed by \citet{FulleEtal2019}.
\emph{(c)} and \emph{(d)} could quench dust activity in fallback regions, but do not play a role in consolidated areas. 

For the coma-related effects (\emph{(e)} and \emph{(f)}), it is worth repeating that the observed dust brightness (or equivalently $Af\rho$) is proportional to the dust loss rate only if the dust velocity, size distribution, and composition do not change.
A variation of dust brightness might therefore \emph{in principle} be interpreted as a change in dust loss rate or as a change of any of the parameters above.
\emph{(e)} We rule out significant differences in the dust properties in different areas (i.e., at different times) due to composition variation, as none was observed in VIRTIS coma observations \citep{Rinaldi2016}.
\emph{(f)} If the coma would be dominated by larger particles, they would tend to reduce the observed dust brightness for the same dust production rate in kg/s, due to the smaller reflecting area of fewer but larger particles.
This would play a role in fallback regions, but not in the consolidated regions that we are considering. Measurements of the dust size distribution from Rosetta's in situ instruments are difficult to separate by different surface areas as they were measured over extended periods, during which the sub-spacecraft location changed considerably, but there are hints of variation. For example, GIADA detected more compact particles from Hapi (a fallback region) and more `fluffy' aggregates elsewhere \citep{DellaCorte2015}, suggesting variations in the size distribution, but again this does not demonstrate differences between different consolidated terrains. Differences were observed in size distribution in the dust released during outbursts, relative to the background coma, but this may be related to the (poorly understood) outburst process rather than regional differences \citep{BockeleeMorvan2017}.

In summary, the best explanation of our observations in the red box in Fig. \ref{fig:illumination_map} is a quenched dust activity due to high cohesion of surface material typical of consolidated regions.

The observations in the green box in Fig. \ref{fig:illumination_map}, an increased water activity in the Southern regions with respect to model expectations, are likely to be attributed to regional changes of volatile content or access to these.
Our simple thermal model assumes the same thickness of the desiccated layer for the Northern and Southern hemisphere.
Due to higher erosion rates in the South, this is likely not the case, and a shallower desiccated layer, or a larger area fraction of ice \citep{FougereEtAl2016}, would explain our observations.

A more thorough investigation in the near future demands application of more sophisticated thermo-physical models to treat not only non-uniform properties of the nucleus subsurface but also activity of multiple volatile species.
It would be necessary to resolve the temperatures in layers deeper than the diurnal skin depth as considered in the current study.
A more detailed characterisation of the physical processes in the subsurface, such as phase change and mass transfer of volatiles as well as the resulting material loss, is also desired, as the phenomena already proved to strongly influence the energy budget of the system \citep{deSanctisEtAl1999, CapriaEtAl2000, PrialnikEtAl2004, GortsasEtAl2011}.

\section{Summary and conclusion}
\label{sec:conclusion}
We have analysed one OSIRIS and one VIRTIS-M data set acquired on 27 April 2015, when the comet was at 1.76 au from the Sun in the inbound arc. 

No strong temporal correlation between total dust brightness and water production rates is found, despite water being still the main driver of dust activity at this period in time.
The observed increased water activity in the Southern regions with strong seasonal variations, with respect to model expectations, is likely to be attributed to regional changes of volatile content or access to this.
The best explanation for the drastic decrease in dust brightness when consolidated regions are illuminated is a quenched dust activity due to the high cohesion of surface material.
These observations show that, when 67P is approaching perihelion, the dust activity cannot be understood based on water-driven activity alone.
This is in agreement with other modelling results on the seasonal evolution of the near-nucleus coma, which show that the correlation observed earlier in the mission, between the observed dust coma and a modelled water coma from a homogeneously sublimating nucleus, is significantly degraded \citep{ShiEtAl2018EPSC}. 

\begin{acknowledgements}
OSIRIS was built by a consortium led by the Max-Planck-Institut f\"ur Sonnensystemforschung, G\"ottingen, Germany, in collaboration with CISAS, University of Padova, Italy, the Laboratoire d'Astrophysique de Marseille, France, the Instituto de Astrof\'isica de Andalucia, CSIC, Granada, Spain, the Scientific Support Office of the European Space Agency, Noordwijk, The Netherlands, the Instituto Nacional de T\'ecnica Aeroespacial, Madrid, Spain, the Universidad Polit\'echnica de Madrid, Spain, the Department of Physics and Astronomy of Uppsala University, Sweden, and the Institut  f\"ur Datentechnik und Kommunikationsnetze der Technischen Universit\"at  Braunschweig, Germany.
The support of the national funding agencies of Germany (DLR), France (CNES), Italy (ASI), Spain (MEC), Sweden (SNSB), and the ESA Technical Directorate is gratefully acknowledged.
We thank the Rosetta Science Ground Segment at ESAC, the Rosetta Mission Operations Centre at ESOC and the Rosetta Project at ESTEC for their outstanding work enabling the science return of the Rosetta Mission. 
\newline
VIRTIS was built by a consortium, which includes Italy, France, and Germany, under the scientific responsibility of the Istituto di Astrofisica e Planetologia Spaziali of INAF, Italy, which also guides the scientific operations. The VIRTIS instrument development, led by the prime contractor Leonardo-Finmeccanica (Florence, Italy), has been funded and managed by ASI, with contributions from Observatoire de Meudon financed by CNES, and from DLR. We thank the Rosetta Science Ground Segment and the Rosetta Mission Operations Centre for their support throughout all the phases of the mission. The VIRTIS calibrated data will be available through the ESA's Planetary Science Archive Website (www.rssd.esa.int) and is available upon request until posted to the archive. We thank the following institutions and agencies for support of this work: Italian Space Agency (ASI, Italy) contract number I/024/12/1, Centre National d'\'Etudes Spatiales (CNES, France), DLR (Germany), NASA (USA) Rosetta Program, and Science and Technology Facilities Council (UK).
\end{acknowledgements}

\bibliographystyle{aa}
\bibliography{bibliography,bibliography_VR}
%
%

\begin{appendix} 
\section{Dust and gas intensities, column densities and production rates}

\begin{table*}[t]
\caption{Dust and gas measured in VIRTIS-M and OSIRIS data.}
\label{tab:dust_gas_fraction}
\footnotesize\centering
\begin{tabular}{lcccccccccc}
\hline
\hline
\\
\# & $A_{\pm45\degr}$ & $A_{360\degr}$ & Af$\rho$ [m]& $F_{\pm45\degr}$ [\%]& $F_{\pm90\degr, D}$ [\%]& $F_{\pm90\degr, N}$ [\%] & $\phi$ [\degr]\\
\\
\hline \hline
\\
H$_2$O&10$^{-5}$ W m$^{-2}$sr$^{-1}$ & 10$^{-5}$ W m$^{-2}$sr$^{-1}$ &&&&&
\\
\hline
\\
1&(10.0 $\pm$ 0.2) & (4.0 $\pm$ 0.2)& -- & 56 & 78 & 22 & 132.9 \\
2&(11.1 $\pm$ 0.2)& (4.1 $\pm$ 0.2)& -- & 60 & 83 & 17 & 108.3 \\
3&(12.4 $\pm$ 0.2) &(5.4 $\pm$ 0.2) & -- & 50 & 77 & 23 & 81.7 \\
4&(10.6 $\pm$ 0.2)&(5.1 $\pm$ 0.2) & -- & 48 & 73 & 27 & 52.8 \\
5& (10.5 $\pm$ 0.2) &(4.9 $\pm$ 0.2)& -- & 48 & 78 & 22 & 355.9 \\
6&(11.4 $\pm$ 0.2)& (4.8 $\pm$ 0.2)& -- & 51 &	77 & 27 & 311.6 \\
7& (11.1 $\pm$ 0.2)& (4.9 $\pm$ 0.2) & -- & 52 &	78 & 22 & 268.4 \\
8&(11.0 $\pm$ 0.2)  & -- & -- & -- &	-- & -- & 222.9 \\
9& -- & -- & -- & --   &--    & --   & 216.6 \\
\hline
\\
CO$_2$&10$^{-6}$ W m$^{-2}$sr$^{-1}$ & 10$^{-6}$ W m$^{-2}$sr$^{-1}$ &&&&&
\\
\hline
\\
1&	(4.8 $\pm$ 2.0) &(6.7 $\pm$ 2.0) & -- & 19 & 49 & 51 & 132.9 \\
2&(4.7 $\pm$ 2.0) &(6.6 $\pm$ 2.0) & -- & 19 & 47 & 53 & 108.3 \\
3&(4.8 $\pm$ 2.0)&(7.4 $\pm$ 2.0)& -- & 16 & 52 & 48 & 81.7 \\
4& (5.3 $\pm$ 2.0)&(7.1 $\pm$ 2.0) & -- & 20 &	54 & 46 & 52.8 \\
5& (10.6 $\pm$ 2.0)&(8.7 $\pm$ 2.0)  & -- & 38 & 66 & 34 & 355.9 \\
6& (12.9 $\pm$ 2.0)&(8.9 $\pm$ 2.0) & -- & 38 & 52 & 48 & 311.6 \\
7& (10.5 $\pm$ 2.0) & (9.1 $\pm$ 2.0) & -- & 33 & 58 & 42 & 268.4 \\
8& (5.2 $\pm$ 2.0)& -- & -- & -- & -- & -- & 222.9 \\
9& -- & -- & -- & --   & --   & --   & 216.6 \\
\hline
\\
Dust&(10$^{-6}$ SI)&(10$^{-6}$ SI)&&&&&&&
\\
\hline
\\
1 & 13.00 $\pm$ 0.35 & -- & --& -- & --& --& 132.9 \\  
2 & 13.18 $\pm$ 0.27& --  & --& -- & --& --& 108.3 \\
3 & 15.10 $\pm$ 0.20 & --  & --& -- & --& --& 81.7 \\
4 & 16.21 $\pm$  0.26 & -- & --& -- & --& --& 52.8 \\
5 & 20.49 $\pm$ 0.33 & --  & --& -- & --& --& 355.9 \\
6 & 18.14 $\pm$ 0.20 & -- & --& -- & --& --& 311.6 \\
7 & 16.87 $\pm$ 0.37 & -- & --& -- & --& --& 268.4 \\
8 & 16.17 $\pm$ 0.32 & -- & --& -- & --& --& 222.9 \\
9 & 16.34 $\pm$ 0.45 &  --  & --& -- & --& --& 216.6 \\
\hline
\\
a& 16.38 $\pm$ 0.04 & 8.98 $\pm$ 0.02& 1.28&46 & 70  & 30  & 249.7 \\
b& 15.66 $\pm$ 0.04 & 8.36 $\pm$ 0.02& 1.19&47 & 68  & 32  & 220.7\\
c& 15.52 $\pm$ 0.04 & 8.39 $\pm$ 0.02& 1.19&46 & 70  & 30  & 191.3 \\
d& 13.46 $\pm$ 0.04 & 7.58 $\pm$ 0.02& 1.08&44 & 67  & 33  & 143.6 \\
e& 13.39 $\pm$ 0.04 & 7.30 $\pm$ 0.02& 1.04&46 & 68  & 32  & 117.1 \\
f& 15.78 $\pm$ 0.05 & 7.50 $\pm$ 0.02& 1.07&53 & 73  & 27  & 88.1 \\
g& 16.51 $\pm$ 0.05 & 7.60 $\pm$ 0.02& 1.08&54 & 76  & 24  & 59.2 \\
h& 16.78 $\pm$ 0.05 & 7.84 $\pm$ 0.02& 1.12&53 & 77  & 23  & 44.7 \\
i& 18.92 $\pm$ 0.05 & 9.22 $\pm$ 0.02& 1.31&51 & 75  & 25  & 15.3 \\
j& 18.95 $\pm$ 0.05 & 9.57 $\pm$ 0.02& 1.36&49 & 74  & 26  & 352.1 \\
k& 18.04 $\pm$ 0.05 & 8.91 $\pm$ 0.02& 1.27&51 & 77  & 23  & 323.2 \\
l& 18.22 $\pm$ 0.05 & 8.75 $\pm$ 0.02& 1.25&52 & 75  & 25  & 294.2 \\
m& 17.33 $\pm$ 0.05 & 8.62 $\pm$ 0.02& 1.23&50 & 75  & 25  & 282.2 \\
n& 15.22 $\pm$ 0.05 & 7.97 $\pm$ 0.02& 1.13&48 & 70  & 30  & 229.1 \\
\hline
\end{tabular}
\flushleft
{\bf{Note}}:
{\it{Column 1}}: Assigned number and letter for each image.
{\it{Column 2}}: Average flux in the $\pm$ 45\degr mask (in W m$^{-2}$ sr$^{-1}$ nm$^{-1}$ for the dust).
{\it{Column 3}}: H$_2$ and CO$_2$: Average emitted band intensity calculated inside the annulus. Dust: Average flux in the full annulus (in W m$^{-2}$ sr$^{-1}$ nm$^{-1}$ nm).
{\it{Column 4}}: Af$\rho$.
{\it{Column 5}}, {\it{Column 6}} and {\it{Column 7}}: fraction of flux in the $\pm$ 45\degr, $\pm$ ~90\degr \; (subsolar) and $\pm$ ~90\degr \;(antisolar) masks, respectively, compared to the full annulus (in \%).
{\it{Column 8}}: Subsolar longitude (in \degr).
\end{table*}

\begin{table*}[t]
\caption{Gas column densities and production rates.}
\label{tab:prod_rate}
\footnotesize\centering
\begin{tabular}{lcccccccc}
\hline
\hline
\\
\# & $v_\text{out}$ [m/s]  & $g_0$ [W molec.$^{-1}$]& $A_{360\degr}$ [W m$^{-2}$ sr$^{-1}$]  & $n_{360\degr}$ [molec. $m^{-2}$]   & $Q$ [molec. $s^{-1}$] & $Q$ [kg $s^{-1}$] &$\phi$ [\degr] \\
\\
\hline \hline
\\
H$_2$O
\\
\hline
\\
1&580& 2.745$\times$10$^{-23}$ & (3.96 $\pm$ 0.20) $\times$10$^{-5}$ &(5.65 $\pm$ 0.29) $\times$10$^{19}$   & (4.06 $\pm$ 0.21) $\times$10$^{26}$  &12.16 $\pm$ 0.61 &132.9\\
2& 580&2.745$\times$10$^{-23}$ &(4.14 $\pm$ 0.20) $\times$10$^{-5}$  & (5.90 $\pm$ 0.29) $\times$10$^{19}$&(4.24 $\pm$ 0.21) $\times$10$^{26}$   &12.70 $\pm$ 0.61 & 108.4\\
3&580& 2.745$\times$10$^{-23}$& (5.36 $\pm$ 0.20) $\times$10$^{-5}$& (7.63 $\pm$ 0.29) $\times$10$^{19}$  &(5.49 $\pm$ 0.20) $\times$10$^{26}$   &16.41 $\pm$ 0.61   &  81.7\\
4& 580&2.745$\times$10$^{-23}$& (5.07 $\pm$ 0.20) $\times$10$^{-5}$& (7.22 $\pm$ 0.29) $\times$10$^{19}$   &(5.19 $\pm$ 0.20) $\times$10$^{26}$   &15.53 $\pm$ 0.61  &52.8\\
5&580 & 2.745$\times$10$^{-23}$ & (4.88 $\pm$ 0.20) $\times$10$^{-5}$ & (6.95 $\pm$ 0.29) $\times$10$^{19}$ &(5.00 $\pm$ 0.20) $\times$10$^{26}$   &14.95 $\pm$ 0.61 &355.9\\
6& 580& 2.745$\times$ 10$^{-23}$& (4.77 $\pm$ 0.20) $\times$10$^{-5}$ & (6.78 $\pm$ 0.29) $\times$10$^{19}$&(4.88 $\pm$ 0.20) $\times$10$^{26}$   &14.60 $\pm$ 0.61 &311.6\\
7& 580&2.745$\times$ 10$^{-23}$ &(4.91 $\pm$ 0.20) $\times$10$^{-5}$ & (6.99 $\pm$ 0.29) $\times$10$^{19}$&(5.03 $\pm$ 0.20) $\times$10$^{26}$ &15.03 $\pm$ 0.61 & 268.4 \\
\hline
\\ 
CO$_2$    
\\
\hline
\\
1&380& 1.25$\times$10$^{-22}$ &(6.7 $\pm$ 2.0) $\times$10$^{-6}$ &(2.11 $\pm$ 0.63) $\times$10$^{18}$   & (0.99 $\pm$ 0.30) $\times$10$^{25}$  & 0.73 $\pm$ 0.22 &132.9\\
2& 380&1.25$\times$10$^{-22}$ &(6.6 $\pm$ 2.0) $\times$10$^{-6}$  & (2.08 $\pm$ 0.63) $\times$10$^{18}$& (0.98 $\pm$ 0.30) $\times$10$^{25}$   & 0.72 $\pm$ 0.22& 108.3\\
3&380& 1.25$\times$10$^{-22}$& (7.4 $\pm$ 2.0) $\times$10$^{-6}$& (2.33 $\pm$ 0.63) $\times$10$^{18}$  & (1.10 $\pm$ 0.29) $\times$10$^{25}$  &  0.80 $\pm$ 0.22   &  81.7  \\
4& 380&1.25$\times$10$^{-22}$&(7.1 $\pm$ 2.0) $\times$10$^{-6}$& (2.23 $\pm$ 0.63) $\times$10$^{18}$   & (1.05 $\pm$ 0.29) $\times$10$^{25}$   &0.77 $\pm$ 0.22&52.8   \\
5&380 & 1.25$\times$10$^{-22}$ & (8.7 $\pm$ 2.0) $\times$10$^{-6}$ &(2.71 $\pm$ 0.63) $\times$10$^{18}$ & (1.28 $\pm$ 0.29) $\times$10$^{25}$   & 0.93 $\pm$ 0.22&355.9 \\
6& 380& 1.25$\times$10$^{-22}$& (8.9 $\pm$ 2.0) $\times$10$^{-6}$ &(2.78 $\pm$ 0.63) $\times$10$^{18}$& (1.31 $\pm$ 0.29) $\times$10$^{25}$   &  0.96 $\pm$ 0.22&311.6 \\
7& 380&1.25$\times$10$^{-22}$ &(9.1 $\pm$ 2.0) $\times$10$^{-6}$ &(2.84 $\pm$ 0.63) $\times$10$^{18}$& (1.34 $\pm$ 0.29) $\times$10$^{25}$ & 0.98 $\pm$ 0.22 & 268.4 \\
\hline
\end{tabular}
\flushleft
{\bf{Note}}:
{\it{Column 1}}: Assigned number for each image.
{\it{Column 2}}: Gas outflow speed \citep{Fink2016}.
{\it{Column 3}}: $g$ factor for the H$_2$O and CO$_2$ bands at 1 au.
{\it{Column 4}}: Average emitted band intensity calculated inside the annulus.
{\it{Column 5}}: Average column density calculated inside the annulus.
{\it{Column 6}} and {\it{Column 7}}: H$_2$O and CO$_2$ production rate in molec. s$^{-1}$ and kg s$^{-1}$, respectively.
{\it{Column 8}}: Subsolar longitude.
\end{table*}

\end{appendix}


\end{document}